\documentclass[aps,prd,floats,nofootinbib]{revtex4}
\usepackage{latexsym}
\usepackage{amssymb}
\usepackage[dvips]{graphicx}
\usepackage[applemac]{inputenc}
\usepackage[nottoc]{tocbibind} 
\usepackage{float}
\usepackage{fancyhdr}
\usepackage{amsfonts}
\usepackage{amsmath}
\usepackage{color}
\usepackage{mathrsfs}
\usepackage{bm}
\usepackage{epsfig}
\usepackage{feynmf}
\usepackage{epsfig}

\def\beq{\begin{equation}}
\def\eeq{\end{equation}}
\def\beqa{\begin{eqnarray}}
\def\eeqa{\end{eqnarray}}

\def\paras{\vskip 0.5cm\noindent}

\newcommand{\lsim}{\mbox{\raisebox{-.6ex}{~$\stackrel{<}{\sim}$~}}}
\newcommand{\gsim}{\mbox{\raisebox{-.6ex}{~$\stackrel{>}{\sim}$~}}}

\def\erf{\rm erf}
\def\half{\frac{1}{2}}

\def\threehalf{\frac{3}{2}} 

\def\cm{\,{\rm cm}}
\def\km{\,{\rm km}}

\def\s{\,{\rm s}}

\def\kmps{\km\s^{-1}}
\def\pc{\,{\rm pc}}
\def\kpc{\,{\rm kpc}}

\def\kev{\,{\rm keV}}

\def\gev{\,{\rm GeV}}

\def\G{\,{\rm G}}
\def\pb{\,{\rm pb}}

\def\umin{u_{\rm min}}
\def\vmax{v_{\rm max}}
\def\umax{u_{\rm max}}
\def\vescp{v_{\rm escp}}
 
\def\msun{\,M_\odot}
\def\rsun{\,R_\odot}
\def\rt{r_t}
\def\vsun{\,v_\odot}
\def\vcsun{v_{c,\odot}}
\def\vE{v_{\rm E}}
\def\vEorb{v_{\rm E, orb}}
\def\ftilde{\tilde{f}}
\def\mchi{m_\chi}
\def\boldx{{\bf x}}
\def\boldv{{\bf v}}
\def\boldvE{\boldv_{\rm E}}
\def\boldu{{\bf u}}

\def\mathcalR{{\mathcal R}}
\def\mathcalC{{\mathcal C}}
\def\mathcalE{{\mathcal E}}
\def\mathcalM{{\mathcal M}}
\def\mathcalK{{\mathcal K}}
\def\rhodm{\rho_{\rm\scriptscriptstyle DM}}

\def\rhovis{\rho_{\rm\scriptscriptstyle vis}}
\def\rhochi{\rho_\chi}

\def\rhodmsun{\rho_{{\rm\scriptscriptstyle DM},\odot}}
\def\Phidm{\Phi_{\rm\scriptscriptstyle DM}}
\def\Phivis{\Phi_{\rm\scriptscriptstyle vis}}
\def\vdispsq{\langle v^2 \rangle}
\def\vdisp{\langle v^2 \rangle^{1/2}}
\def\vdispdm{\vdisp_{\rm\scriptscriptstyle DM}}

\begin{document}
\title{\bf 
Direct detection of WIMPs~:~Implications of a self-consistent 
truncated isothermal model of the Milky Way's dark matter halo 
} 
\author{Soumini Chaudhury$^a$\footnote{soumini.chaudhury@saha.ac.in}, 
Pijushpani Bhattacharjee$^a$\footnote{pijush.bhattacharjee@saha.ac.in},
Ramanath Cowsik$^b$\footnote{cowsik@wuphys.wustl.edu}
}
\affiliation{
$^a$ AstroParticle Physics and Cosmology Division 
\& Centre for AstroParticle Physics (CAPP),
Saha Institute of Nuclear Physics, 1/AF Bidhannagar,
Kolkata 700064. India \\
$^b$ McDonnell Center for the Space Sciences and Department of
Physics, Washington University, St.~Louis, MO 63130, USA
}
\begin{abstract}
\noindent 
Direct detection of Weakly Interacting Massive Particle 
(WIMP) candidates of Dark Matter (DM) is studied within the 
context of a self-consistent truncated isothermal model of the 
finite-size dark halo of the Galaxy. The halo model, based on the 
``King model" of the phase space distribution function of collisionless 
DM particles, takes into account the modifications of the phase-space 
structure of the halo due to the gravitational influence of the observed 
visible matter in a self-consistent manner. The parameters of the halo 
model are determined by a fit to a recently determined circular rotation 
curve of the Galaxy that extends up to $\sim$ 60 kpc. Unlike in the 
Standard Halo 
Model (SHM) customarily used in the analysis of the results of WIMP 
direct detection experiments, the velocity 
distribution of the WIMPs in our model is non-Maxwellian with a cut-off 
at a maximum velocity that is self-consistently determined by the model 
itself. For our halo model that provides the best fit to the  
rotation curve data, the 90\% C.L. upper limit on the 
WIMP-nucleon spin-independent cross section from the recent results of 
the CDMS-II experiment, for example, is $\sim 5.3\times10^{-8}\pb$ at a 
WIMP mass of $\sim$ 71 GeV. We also find, using the original 2-bin 
annual modulation amplitude data on the nuclear recoil event rate seen   
in the DAMA experiment, that there exists a range of small WIMP 
masses, typically $\sim$ 2 -- 16 GeV, within which DAMA collaboration's 
claimed annual modulation signal purportedly due to WIMPs is compatible  
with the null results of other experiments. These results, based 
as they are on a self-consistent model of the dark matter halo of the 
Galaxy, strengthen the possibility of low-mass ($\lsim 10\gev$) WIMPs as 
a candidate for dark matter as indicated by several earlier 
studies performed within the context of the SHM. A more rigorous 
analysis using DAMA bins over smaller intervals should be able to better 
constrain the ``DAMA regions" in the WIMP parameter space 
within the context of our model. 
\end{abstract}
\maketitle
\newpage
\section{Introduction
\label{sec:intro}}
\noindent 
One of the most favored candidates for the Dark Matter (DM), that 
constitutes roughly a quarter of the mass-energy density of the 
Universe~\cite{wmap2010},   
is some kind of Weakly Interacting Massive Particles (WIMPs) 
with masses typically between a few GeV and a few TeV. Such particles 
are predicted in many models of physics beyond the Standard Model of 
particle physics such as
models involving supersymmetry or higher (spatial) dimensional
theories~\cite{Jungman_etal_PhysRep_96,
Bertone_etal_PhysRep_05,Hooper_Profumo_KKDM_PhysRep_07}. Owing to
their weak interactions, such particles would ``freeze out" in
the early Universe at a characteristic temperature scale of $\sim$ MeV,
and would be present in the Universe today as `cold' dark matter with a 
calculable abundance 
that can naturally explain the measured average cosmological density of  
dark matter in the Universe. Also, the measured rotation curves of 
spiral galaxies, such as that of our own, the Milky Way, out to and 
beyond their visible edges, can be explained naturally under the 
hypothesis that the visible galaxies are embedded in much larger, 
roughly spherical, dark matter ``halos''. The true extents and 
the masses of such halos are, however, unknown.

\paras
Following early suggestions~\cite{Goodman_Witten_85,DFS_86,FFG_88} (see 
\cite{Lewin_Smith_96} for a review), several 
experiments worldwide are currently engaged in attempts to directly 
detect these hypothetical WIMP DM particles comprising the DM Halo 
of the Milky Way (the Galaxy, hereafter) in terrestrial laboratories. 
These direct detection experiments attempt to detect nuclear recoils 
resulting from interaction of WIMPs with a suitably chosen target 
material in underground laboratory detectors with low backgrounds. 

\paras 
The DAMA/NaI~\cite{dama_NaI} and DAMA/LIBRA~\cite{dama_libra_08}
experiments (hereafter together simply referred to as the ``DAMA"
collaboration) have reported an annual modulation signal in their event
rate at a confidence level of $8.2\sigma\,$ based on their combined
data~\cite{dama_libra_08}. They interpret this as evidence for WIMPs,
ascribing the modulation to the periodic variation of the flux of
WIMPs passing through the detector on earth caused by the orbital motion
of the earth around the sun. The CDMS-II  
collaboration~\cite{cdms-II_Ge_PRL_10} has also reported  
two candidate events in their ``signal region" with an estimated    
probability of $\sim$ 23\% of observing 2 or more background 
events in that region, for the estimated level of the background. 
Several other experiments such as CRESST~\cite{cresst}, 
XENON10~\cite{xenon10a,xenon10b}, COUPP~\cite{coupp} and 
PICASSO~\cite{picasso}, for example, have reported only null results so 
far. These results provide important constraints 
on the WIMP mass and WIMP-nucleon interaction cross section. The 
question of compatibility of the DAMA results~\cite{dama_libra_08} 
with the negative results from other experiments has also been the 
subject of several recent 
studies which include the earlier overlooked effect of ``ion channeling" 
(see below) on the analysis of the DAMA results   
~\cite{dama_compatibility_others,Petriello-Zurek_08,Savage_etal_08}. 

\paras
In order to derive constraints on (or to determine) the mass and
interaction cross section of the unknown WIMP DM particles
from the results of these direct detection experiments, one requires the
density and velocity distribution of the WIMPs in the
solar neighborhood as crucial input parameters. These parameters are
{\it a priori} unknown. In the customary analysis the DM halo of the
Galaxy is assumed to be described by a single component
isothermal sphere~\cite{Binney_Tremaine} with a Maxwellian velocity
distribution in the Galactic rest frame given by
\beq
f(\boldx,\boldv)d^3\boldv=4\pi
\rho(\boldx)\left[\frac{3}{2\pi\vdispsq}\right]^{\threehalf}v^2
\exp\left[-\threehalf\frac{v^2}{\vdispsq}\right]dv\,.
\label{eq:maxwellian}
\eeq
Here $v=|\boldv|$, \  $\rho(\boldx)$ is the
density at the location $\boldx$, and $\vdisp$ is the velocity
dispersion. The justification for this Maxwellian form comes from the
work of Lynden-Bell~\cite{LyndenBell_67} who argued that evolution of
systems of collisionless particles (such as WIMPs) under the process of
gravitational collapse is governed by the process of
``violent relaxation'' whereby the observationally relevant
`coarse-grained' phase space distribution function (DF) of the system 
rapidly
relaxes, due to collective effects, to a quasi-Maxwellian stationary
state which is a steady-state solution of the collisionless
Boltzmann equation.\footnote{The `fine-grained' DF may, however, never
reach equilibrium. }

\paras
The DM density in the solar neighborhood is usually taken to be
$\rhodmsun\sim0.3\pm0.1\gev/\cm^3$, following the analysis of the
observed velocity distribution of stars transverse to the Galactic disk
near the sun, as suggested by Oort~\cite{Oort} and extended by
Bahcall~\cite{Bahcall_84}. A recent ``model-independent" analytical
study~\cite{Salucci_etal_1003_3101} quotes
$\rhodmsun=0.43\pm0.11\pm0.10\gev/\cm^3$.
Also, a recent attempt to directly extract this density from the results
of cosmological large N-body simulations yields a value of
$\rhodmsun\sim0.37\gev\cm^{-3}$~\cite{Ling_etal_simulations_09}. The
last Reference also finds a DM particle velocity distribution that can
in general be called ``quasi-Maxwellian", albeit with significant
deviation from the pure Maxwellian form especially at the high velocity
end where it drops off more sharply compared to the pure Maxwellian.

\paras
The value of the velocity dispersion, $\vdisp$, the single parameter
that characterizes the Maxwellian velocity distribution
(\ref{eq:maxwellian}) of the DM particles, is usually determined from
the relation~\cite{Binney_Tremaine},
${\vdisp}=\sqrt{\threehalf} v_{c,\infty}$, between the velocity
dispersion of the particles constituting a single-component
self-gravitating isothermal sphere and the
asymptotic value of the circular rotation speed, $v_{c,\infty}$,
of a test particle in the gravitational field of the isothermal sphere.
Neglecting the effect of the visible matter on the DM halo, and assuming
$v_{c,\infty}\approx v_{c,\odot}\approx 220\kmps$, where $v_{c,\odot}$
is the measured value of the circular rotation velocity of the Galaxy in
the solar neighborhood, one gets
${\vdisp}_{\rm iso}\approx270\kmps$.~\footnote{A recent study suggests a
somewhat higher value of
$v_{c,\odot}\approx 250\kmps$~\cite{new_rot_speed}, which would imply a
correspondingly
higher value of ${\vdisp}_{\rm iso}\approx306\kmps$.}
This isothermal sphere model of the DM halo with a value of the DM
velocity dispersion $\vdispdm\simeq270\kmps$ and local value of the DM
density $\rhodmsun\simeq0.3\gev\cm^{-1}$ is what is often
referred to as the {\it Standard Halo Model} (SHM). This SHM is
generally taken as a sort of benchmark model of
the DM halo of the Galaxy, and the results of various WIMP DM search
experiments are mostly analyzed within the context of this SHM.

\paras
Whereas the SHM would suffice for the initial design of the experiments,
it is necessary to consider improved models to analyze the results of
the experiments, which are becoming progressively more sophisticated.
The reasons for this are many: (1)~It is well-known that the isothermal
sphere has a mass that linearly increases with its radius $r$ and tends
to $\infty$ as $r\to \infty$. Clearly, such a system cannot 
represent a realistic DM halo of finite physical size. (2)~The 
rotational speed in the isothermal sphere model increases linearly for 
small $r$, has an
oscillatory dependence on $r$ for intermediate $r$, and tends to a
constant only asymptotically as $r\to\infty$. Thus there is no
strict reason why $v_{c,\infty}$ should be equal to $v_{c,\odot}$.
(3)~More importantly, the presence of the visible matter further
complicates the situation; it contributes significantly to balancing the
centrifugal forces of rotation at the solar circle. The density profile
of an isothermal sphere is obtained as the solution to a second-order
non-linear differential equation with appropriate boundary conditions at
the origin. Thus the presence of a significant amount of visible matter
within the solar circle will substantially change the nature of the
solutions. (4)~It is a
common practice, in the context of the phenomenology of DM direct
detection experiments, to truncate the speed distribution
(\ref{eq:maxwellian}) at some chosen value of the local (solar
neighborhood) escape speed $\vescp$; see, e.g.,
\cite{FFG_88,Lewin_Smith_96}). However, this is not in general a
self-consistent procedure because the resulting ``truncated Maxwellian"
speed distribution will not necessarily be a solution of the
steady-state collisionless Boltzmann equation appropriate for a finite
physical system. In addition, since the rotation curve for such a
truncated Maxwellian will not in general be asymptotically flat, the
relation ${\vdisp}=\sqrt{\threehalf} v_{c,\infty}$ used to determine the
value of $\vdisp$ in the Maxwellian speed distribution of the
isothermal sphere, as done in the SHM, will not be valid. However, there
are indeed well understood procedures that allow truncation of the DM
halo at a finite radius and the velocity distribution at appropriate
limits in a self-consistent manner.

\paras
In this paper we construct a self-consistent model of the phase-space
structure of the finite-size DM halo of the Galaxy, and study its
implications for the analysis of the results of the direct detection
experiments. The model is based
on describing the phase-space DF of the DM particles in the halo by the
so-called ``lowered" (or truncated) isothermal models (often called
``King models")~\cite{Binney_Tremaine}, which are proper self-consistent
solutions of the collisionless Boltzmann equation representing nearly
isothermal systems of finite physical size and mass. At every location
$\boldx$ within the system a DM particle can have speeds up to a maximum
speed $\vmax(\boldx)$, which is self-consistently determined by the 
model itself. A particle of velocity $\vmax(\boldx)$ at $\boldx$ within 
the system can just reach its outer boundary, generally called the
truncation radius, where the DM density by construction vanishes. Note 
that this maximum speed $\vmax$ is not the same as the escape speed 
$\vescp$, which is defined as the speed required for a particle to 
escape from the given location out to infinity. The escape speed is 
always larger than the maximum speed. For a King model DM halo of finite 
size, it is the maximum speed and not the escape speed that is relevant 
in considerations of direct detection experiments. As we shall see 
below, the speed distribution of
the particles constituting a lowered isothermal model can be described
as ``quasi-Maxwellian" not unlike what is seen in recent
simulations~\cite{Ling_etal_simulations_09}. Apart from taking into 
account the finite size of the DM halo, we also explicitly account for 
the mutual gravitational interaction between the visible matter and the 
DM, as this interaction influences both the density profile and the 
velocity distribution of the dark matter particles. When the 
visible matter density is set to zero and the truncation radius is set 
to infinity, our halo model becomes identical to that of a 
single-component isothermal sphere used in the SHM. 

\paras
The model we develop here is along the lines described in earlier 
formulations~\cite{CRB_PRL_96&97,CRBM_NewAstron_07} which included the 
effects of the gravitation of the visible matter on the dark matter in 
a self-consistent manner.  
Stated simply, these models account for the fact that the DM particles 
move in a gravitational potential to which both the visible matter and 
the dark matter make contributions. As the distribution of 
the visible matter may be fixed from observations, the parameters of the 
self-consistent velocity distribution function of the DM particles in 
presence of the visible matter can be determined by comparing the
theoretically derived rotation curve with the observed rotation curve 
of the Galaxy. The velocity distribution function so 
determined is then used in the calculation of the direct detection 
rates. 

\paras 
Of particular interest in this context is the rotation curve 
recently reported by 
Xue et al~\cite{rc_60kpc_Xue_etal_08}, which extends up to a galactocentric 
distance of $\sim60\kpc$. This rotation curve is derived from the 
kinematics of a sample of $\sim$ 2400 blue horizontal-branch (BHB)
stars taken from the Sloan Digital Sky Survey (SDSS) DR6~\cite{SDSS_DR6}
database. A noticeable feature of this rotation curve is that it 
gently falls from the adopted value at the Sun's location,  
$\vcsun\approx220\kmps$, to $\sim180\kmps$ at $\sim60\kpc$. It is of 
considerable interest to study the implications of such a  
rotation curve for the direct detection experiments. In this paper, 
therefore, we use this new rotation curve for determining the 
parameters of our halo model.  

\paras 
For the isothermal sphere model of the DM halo, the main effect of the 
gravitational influence of the visible matter on the structure of the DM 
halo is to shrink its `core radius'~\cite{Binney_Tremaine},  
$r_0\equiv \big(3\vdispsq/(4\pi G\rho_0)\big)^{1/2}$, 
where $\rho_0$ is the central density~\footnote{The 
gravitational effect of the Galactic disk, of course, flattens the 
isothermal `sphere'. Our formalism in this paper automatically 
and self-consistently incorporates this effect. Below, unless otherwise 
specified, by core radius we refer to that of the 
density distribution in the plane of 
the Galactic disk. Note, however, that while the resulting density 
distribution of the halo becomes anisotropic, one can still  
assume, as we do in this paper, that the velocity distribution of the DM 
particles remains isotropic everywhere, although anisotropic 
(velocity) models are also possible~\cite{Binney_Tremaine}. The 
effects of anisotropic and also triaxial halo models on the event 
rates in direct detection experiments and the resulting nature of the 
exclusion limits on WIMP parameters have been studied earlier; see, 
e.g., Ref.~\cite{Anne_Green_PRD_02} and references therein.}.  
This is because the DM is ``pulled in" by the visible matter, increasing 
its effective central density. Thus the characteristic core radius of 
an isothermal DM distribution in presence of the visible matter is 
less than that of an isolated isothermal sphere whose properties are 
determined by its own gravity alone. The same, as we show below, is true 
for the lowered isothermal models. A related effect is the 
relative enhancement of the DM density near the plane of the disk due to 
additional gravitational influence of the disk. These effects are 
further discussed in the following section. 

\paras 
We find that, for our best-fit  
halo model with parameter values $\rhodmsun=0.2\gev/\cm^3$ and 
truncation radius $r_t=120\kpc$, that provides the best fit to the 
rotation curve data of Ref.~\cite{rc_60kpc_Xue_etal_08}, 
the null result of the CDMS-II experiment~\cite{cdms-II_Ge_PRL_10}, for 
example, gives a 90\% C.L. upper limit on the WIMP-nucleon 
spin-independent (SI) cross section of $\sim 
5.3\times10^{-8}\pb$ for a WIMP mass of $\sim$ 71 GeV. 
  
\paras
We also study the issue of compatability of the claimed 
positive signal reported by the DAMA collaboration with the null results 
from the other experiments using, for illustration, the DAMA 
collaboration's original 2-bin annual modulation amplitude data. 
We find that for our best-fit halo model, there 
exists a region of the WIMP mass ($\mchi$) vs.~WIMP-nucleon 
spin-independent (SI) cross 
section ($\sigma_{\chi p}^{\rm SI}=\sigma_{\chi n}^{\rm SI}$)
parameter space bounded by $2.6 \lsim\mchi \lsim 10.8\gev$ and 
$1.0\times10^{-2}\gsim \sigma_{\chi p}^{\rm SI}\gsim 
1.2\times10^{-5}\pb$, within which the DAMA's claimed modulation 
signal is compatible with the null results of other experiments. 
Similar ``DAMA-compatible" regions of small WIMP masses obtain also for 
spin-dependent interactions. We, however, recognize that a more rigorous 
analysis using DAMA bins over smaller intervals should be able to better 
constrain the DAMA region within the context of our halo 
model, a task that we will take up in a subsequent work. 

\paras
The rest of the paper is arranged as follows: In section 
\ref{sec:KingModel} we describe our model of the dark halo of the Galaxy 
that includes in a self-consistent manner the gravitational 
effect of the observed visible matter on the structure of dark matter 
halo. The parameters of the model are 
determined by fitting the theoretically calculated rotation curves to 
a recently determined extended rotation curve of the Galaxy up 
to $\sim 60\kpc$. The shapes of the dark matter density and mass 
profiles and the velocity distribution function of the DM particles are 
also derived. The implications of our model for the analysis of the 
results of WIMP direct detection experiments including the 
question of compatibility of DAMA's claimed positive signal with the 
null results of other experiments, are studied in section 
\ref{sec:dd_implications}. Section \ref{sec:summary_conclusions} 
summarizes the main results of this paper and concludes. 

\paras
In this paper we restrict our attention to only elastic scattering of 
WIMPs from nuclei due to spin-independent 
(SI) as well as spin-dependent (SD) WIMP-nucleus interactions. 

\section{The Self-consistent Truncated Isothermal Model of the 
Dark Matter Halo of the Galaxy
\label{sec:KingModel}}
\subsection{The Basic Formalism\label{subsec:KingModel_Basic}}
\noindent 
The Maxwellian velocity distribution (\ref{eq:maxwellian}) used in the 
SHM is simply related to the phase-space DF of the isothermal 
sphere~\cite{Binney_Tremaine},   
\beq
f_{\rm IS}(\boldx,\boldv) = f(E) = 
\frac{\rho_0}{\left(2\pi\sigma^2\right)^\threehalf} 
\exp\left[-E/\sigma^2\right]\,,
\label{eq:iso_DF}
\eeq
with $\vdispsq=3\sigma^2$ and $E=\Phi(\boldx) + \half v^2$, the total 
energy (per unit mass) of the DM particle, $\Phi(\boldx)$ being the 
total gravitational potential with the boundary condition $\Phi(0)=0$, 
so that the density at any point $\boldx$ is   
\beq
\rho_{\rm IS}(\boldx)=\int f_{\rm IS}(\boldx,\boldv) d^3\boldv =     
\rho_0\exp\left[-\Phi(\boldx)/\sigma^2\right]\,,
\label{eq:rhoeq} 
\eeq
where $\rho_0$ is the central density. The isothermal sphere described 
by DF (\ref{eq:iso_DF}) is infinite in extent with a divergent total mass, 
and is characterized by two parameters, $\sigma$ and $\rho_0$.

\paras 
To describe ``nearly isothermal" systems of finite size and 
finite total mass, one must have, in addition to the above two 
parameters of the isothermal sphere, a parameter 
that characterizes the finite size of the system. This is accomplished  
in King models by taking the DF to be of the following  
form~\cite{Binney_Tremaine}: 
\begin{equation}
f_{\rm K}(\boldx,\boldv) \equiv f(\mathcalE)=\left\{ \begin{array}{ll} 
\rho_1
(2\pi\sigma^2)^{-3/2}\left(e^{\mathcalE/\sigma^2} -1\right) &
\mbox{for $\mathcalE > 0$}\,,\\
0 & \mbox{for $\mathcalE\leq 0$}\,,
\end{array}
\right.
\label{eq:king_df}
\end{equation}
where 
\begin{equation}
\mathcalE(\boldx)\equiv\mathcalC - \left(\half v^2 + 
\Phi(\boldx)\right)\equiv 
\Psi(\boldx)-\half v^2\,, 
\label{eq:relative_energy_def}
\end{equation}
is the so-called ``relative energy" and $\Psi(\boldx)\left(\equiv 
-\Phi(\boldx)+\mathcalC\right)$  
the ``relative potential"~\cite{Binney_Tremaine}. The three 
(constant) parameters 
of the system are $\rho_1$ (with dimension of density), $\sigma$ 
(with dimension of velocity) and 
the new parameter $\mathcalC$ (with dimension of potential or 
squared velocity) which is related to the finite 
size of the system (see below). 

\paras
The density at any position $\boldx$ is obtained by integrating 
$f_{\rm K}(\boldx,\boldv)$ over all velocities giving  
\beqa
\rho_{\rm K}(\boldx) & = & 
\frac{\rho_1}{\left(2\pi\sigma^2\right)^{3/2}} 
\int_0^{\sqrt{2\Psi(\boldx)}} dv\, 4\pi v^2 
\left[\exp\left(\frac{\Psi(\boldx)-v^2/2}{\sigma^2}\right) - 
1\right]  \\
 & = & 
\rho_1\left[\exp\left(\frac{\Psi(\boldx)}{\sigma^2}\right)\,
\erf\left(\frac{\surd{\Psi(\boldx)}}{\sigma}\right) - 
\sqrt{\frac{4\Psi(\boldx)}{\pi\sigma^2}}\left(1 + 
\frac{2\Psi(\boldx)}{3\sigma^2}\right)\right]\,,
\label{eq:king_density}
\eeqa
which satisfies the Poisson equation 
\beq
\nabla^2 \Phi(\boldx) = - \nabla^2 \Psi(\boldx) = 4\pi G 
\rho_{\rm K}(\boldx)\,. 
\label{eq:Poisson_1}
\eeq 

\paras
At any location $\boldx$ the maximum speed a particle of the system can 
have is 
\beq 
\vmax (\boldx)=\sqrt{2\Psi(\boldx)}\,,
\label{eq:vmax}
\eeq
at which the relative energy $\mathcalE$ and, as a consequence, the DF 
(\ref{eq:king_df}), vanish.  

\paras 
The King models approach the isothermal sphere solution at small radii 
$r=|\boldx|$, but have 
density profiles that fall off faster than that of the isothermal 
sphere at large 
radii (which makes the total mass finite). Indeed, the density vanishes 
at $r=r_t$, the truncation radius where $\mathcalE=0$, representing the 
outer edge of the system. This is ensured by choosing 
\beq
\mathcalC = \Phi(r_t) \,, \,\,\,\,\,\, 
\mbox{so that} \,\,\, \Psi(r_t)=-\Phi(r_t) + \mathcalC = 0\,.  
\label{eq:rt_mathcalC_reln}
\eeq
As evident from Equation (\ref{eq:rt_mathcalC_reln}), the parameter 
$\mathcalC$ fixes the finite 
size ($r_t$) of the system. Actually, the King models form a sequence, 
each model being parametrized by the value of the ``concentration 
parameter" $C/\sigma^2$. It can be shown~\cite{Binney_Tremaine} that 
the sequence of King models goes over into the isothermal sphere 
in the limit $C/\sigma^2 \to \infty\,$. 

\paras 
Note that the usual escape speed, $\vescp$, defined as 
$\vescp^2(\boldx) = 2\left[\Phi(\infty)-\Phi(\boldx)\right]$ is related 
to the maximum speed $\vmax(\boldx)$ defined by equation 
(\ref{eq:vmax}) through the relation  
\beqa
\vmax^2(\boldx)&=2\Psi(\boldx)=2\left[\Phi(\rt)-\Phi(\boldx)\right]
\nonumber\\
&=\vescp^2(\boldx) - 2\G M(\rt)/\rt\,,
\label{eq:vescp_vmax_reln}
\eeqa
where $M(\rt)$ is the mass contained within $\rt$, with\footnote{Note 
that we have chosen the boundary condition $\Phi(0)=0$, so that 
$\Phi(\infty)$ is a non-zero positive constant.}  
$\G M(\rt)/\rt = \left[\Phi(\infty)-\Phi(\rt)\right]$.  
Note also that, unlike in the case of the (infinite) isothermal sphere, 
the parameter $\sigma$ in the King models is 
{\it not} equal to the one-dimensional velocity dispersion of the 
particles constituting the system; the latter can be
calculated for the DF given above and is a function of $r$, vanishing
at $r=r_t$. In fact, unlike in the case of the Maxwellian DF for which the 
velocity dispersion is linearly related to the most probable speed of the 
particles of the system, there is no simple relation between the velocity 
dispersion and the most probable speed of the particles of the system 
described by a skewed velocity distribution such as that described by the 
King DF (\ref{eq:king_df}), and as such, the velocity dispersion has no 
special 
significance and does not uniquely 
specify the full DF of the King model.  

\paras
In the subsequent discussions we shall take the King model DF 
(\ref{eq:king_df}) as the DF of 
the DM particles constituting the finite-size DM halo of the Galaxy, 
{\it in the rest frame 
of the Galaxy} (the subscript K will be dropped for convenience).     
 
\paras
For a ``pure" DM halo of finite size represented by the DF 
(\ref{eq:king_df}) without including the 
gravitational influence of the visible matter embedded within the 
halo, the gravitational potential $\Phi$ in the above 
equations can be written as       
$\Phi(\boldx)=\Phidm^{(0)}(r)$, where $\Phidm^{(0)}(r)$ is the
spherically symmetric self-gravitational potential of the pure DM halo 
represented by the DF (\ref{eq:king_df}). In reality, however, a
test particle sees the total gravitational potential of both the DM and
the visible matter embedded within the DM halo. Thus, we should write  
\beq
\Phi(\boldx) = \Phidm(\boldx) + \Phivis(\boldx)\,,
\label{eq:Phitotal}
\eeq
where now $\Phidm$ represents the contribution of the DM component to 
the total gravitational potential {\it in presence of the visible 
matter}, and $\Phivis$ is the gravitational potential of the observed 
visible matter of the Galaxy. These potentials satisfy their respective 
Poisson equations  
\beq
\nabla^2\Phidm(\boldx)=4\pi G \rhodm(\boldx)\,,\,\,\,\,\,\,\,\, 
\mbox{and} \,\,\,\, 
\nabla^2\Phivis(\boldx)=4\pi G \rhovis(\boldx)\,, 
\label{eq:poisson_eqns}
\eeq
where the DM density is  
\beq
\rhodm(\boldx)=\rho_1\left[\exp\left(\frac{\Psi(\boldx)}{\sigma^2}\right)\,
\erf\left(\frac{\surd{\Psi(\boldx)}}{\sigma}\right) -
\sqrt{\frac{4\Psi(\boldx)}{\pi\sigma^2}}\left(1 +
\frac{2\Psi(\boldx)}{3\sigma^2}\right)\right]\,,
\label{eq:rhodm_eqn}
\eeq
with 
\beq
\Psi(\boldx)=\Big[\Phidm(|\boldx|=r_t) + \Phivis(|\boldx|=r_t)\Big] 
- \Big[\Phidm(\boldx) + \Phivis(\boldx)\Big]\,,
\label{eq:Psi_eqn}
\eeq
using (\ref{eq:rt_mathcalC_reln}), and $\rhovis$ is the known visible 
matter density of the Galaxy. 

\paras
Note that in presence of the 
visible matter, whose spatial distribution is 
non-spherically symmetric, the DM spatial distribution and hence its 
gravitational potential will also be non-spherically symmetric in 
general. However, for large enough values of the truncation 
radius ($r_t$) of the halo (more specifically, for $r_t$ large compared to 
the visible matter disk's scale length --- the latter being $\sim$ few kpc; 
see below), both $\Phidm$ and $\Phivis$ become close to 
spherically symmetric at $|\boldx| \to r_t$, which we also see in our 
numerical calculations described below. On the other hand, as already 
mentioned in footnote 3 in Section \ref{sec:intro}, the 
velocity distribution will be assumed to remain isotropic. 

\paras
The visible matter distribution $\rhovis(\boldx)$ (and hence the 
potential $\Phivis(\boldx)$) being known from 
various observational data and modeling, solutions of equations 
(\ref{eq:poisson_eqns}), (\ref{eq:rhodm_eqn}), (\ref{eq:Psi_eqn}) with 
appropriate boundary conditions, which we choose as 
\beq
\Phidm(0)=\Phivis(0)=0\,,\,\,\,\,\, {\rm and} \,\,\,\, 
\left(\nabla\Phidm\right)_{|\boldx|=0} = 
\left(\nabla\Phivis\right)_{|\boldx|=0} = 0\,,  
\label{eq:boundary_cond}
\eeq
give us 
a three-parameter family of self-consistent solutions for 
$\rhodm(\boldx)$ and $\Phidm(\boldx)$ for chosen values of the  
parameters ($\rho_1\,, \sigma\,, r_t$). 

\paras
For a given solution, the rotation curve of 
the Galaxy, $v_c(R)$, i.e., the circular rotation velocities in the 
equatorial plane of the Galaxy as a function of the Galactocentric 
distance ($R$) on the equatorial plane, can be calculated from 
\begin{equation}
v_c^2 (R) = R\frac{\partial}{\partial R}\Big[\Phi_{\rm 
total}(R,z=0)\Big] = 
R\frac{\partial}{\partial R}\Big[\Phidm(R,z=0)+\Phivis(R,z=0)\Big]\,,  
\label{eq:v_c_def}
\end{equation}
$z$ being the distance normal to the equatorial plane. The best-fit 
values of the relevant parameters can then be 
determined by comparing the theoretically calculated rotation curves 
with the observed data on the rotation curve of the Galaxy. The 
parameter values so determined are then to be used as inputs to the 
analysis of the results of the direct detection experiments. 

\paras
We have developed an efficient numerical scheme to solve the coupled 
non-linear Poisson equation (\ref{eq:poisson_eqns}) for DM with 
$\rhodm$ given by equations (\ref{eq:rhodm_eqn}) and (\ref{eq:Psi_eqn}) 
with 
boundary conditions (\ref{eq:boundary_cond}), for a given 
potential--density pair of the visible matter, through an 
iterative procedure first discussed in \cite{CRB_PRL_96&97}. Each 
stage of this iteration procedure involves a Poisson solver which is 
tested on a number of exact analytical formulae for potential--density 
pairs given, for example, in \cite{Binney_Tremaine}. The iteration 
process converges typically within ten iterations. We 
have also verified that this iterative scheme yields good 
agreement with the known results for both the non-singular isothermal 
sphere and King Models given in \cite{Binney_Tremaine}, to within a few 
percent. 

\paras
We assume that the density distribution of the visible matter
can be effectively described by the well-known model of a 
spheroidal bulge superposed on an axisymmetric disk~\cite{CO81&KG89}. 
The density distributions of these components are given, respectively, 
by
\begin{equation}
\rho_s(r)=\rho_s(0)\left(1 +\frac{r^2}{a^2}\right)^{-3/2}\,,
\label{eq:rho_sph}
\end{equation}
and
\begin{equation}
\rho_d(r)=\frac {\Sigma_\odot}{2h}e^{-(R-R_0)/R_d} \,\,\, e^{-|z|/h}\,,
\label{eq:rho_disk}
\end{equation}
where $r=(R^2+z^2)^{1/2}\,$. Typical parameter values are 
~\cite{CO81&KG89,Naab_Ostriker_06} $\rho_s(0)=4.2 \times
10^{2}\msun\pc^{-3}\,\,$, $a=0.103\kpc\,$, $R_d=3\kpc\,$, and
$h=0.3\kpc\,$, with $R_0=8.5\kpc\,$, the solar Galactocentric
distance, and $\Sigma_\odot\approx 48\msun\pc^{-2}$, the surface density
of the disk at the solar location. The expressions for the
gravitational potentials, $\phi_s$ and $\phi_d$, corresponding to
above forms of $\rho_s$ and $\rho_d$, are given in \cite{CO81&KG89}. 
Alternatively, they can be directly calculated using our numerical 
Poisson solver. This specifies the $\Phivis = \phi_s + \phi_d$ used in 
the numerical calculations in this paper. 

\paras
There exist more detailed models of the mass distribution of the 
visible matter in the Galaxy than those adopted above, involving 
a stellar ``thick disk" (with a scale height of $\sim 1\kpc$) and the 
disk formed by the Galaxy's interstellar medium (ISM) in 
addition to the stellar spheroid and the ``thin'' disk described by 
equations (\ref{eq:rho_sph}) and (\ref{eq:rho_disk}). 
However, there are also large uncertainties in the values of 
the parameters that characterize these various components, as 
clear from the values of the parameters of the two representative models 
summarized, for example, in Table 2.3 of Ref.~\cite{Binney_Tremaine}. 
We find, as shown below, that the typical values of the various visible 
matter parameters chosen above give reasonably good fit to the rotation 
curve data in the inner Galaxy region, $R\lsim \rsun\approx 8.5\kpc$, 
where the effect of the Dark Matter should be minimal. As such, we 
believe our ``minimal" model of the distribution of the normal matter of 
the Galaxy adopted above is good enough for the purpose of illustrating 
the general nature of effects arising from a self-consistent description 
of the Galaxy's phase space structure that includes the mutual 
gravitational interaction of the visible matter and the Dark Matter 
components of the Galaxy. The formalism of this paper and the numerical 
procedure adopted by us are, however, quite general and applicable to 
any given model of the mass distribution of the Galaxy.       

\paras
In our numerical calculations, we have taken the DM density at the 
solar location, $\rhodmsun=\rhodm(R=R_0,0)$, as the 
``observable" density parameter of the King model instead of the 
parameter $\rho_1$ in equation (\ref{eq:rhodm_eqn}).   
We have run a large number of numerical models of the Galaxy for three 
different values of $\rhodmsun=0.2\,, 0.3\,\,\, \mbox{and}\,\,\, 
0.4\,\gev/\cm^3\,$, in each case with a wide range of values of the 
other two King model parameters $r_t\,$ and $\sigma$ and with the 
visible matter parameters fixed as described above, to determine the 
parameters that best describe the Galaxy's rotation curve data 
mentioned above.  

\subsection{Dark Matter Density Profile
\label{subsec:DM_profiles_etc}}
\begin{figure}[h]
\centering
\begin{tabular}{cc}
\epsfig{file=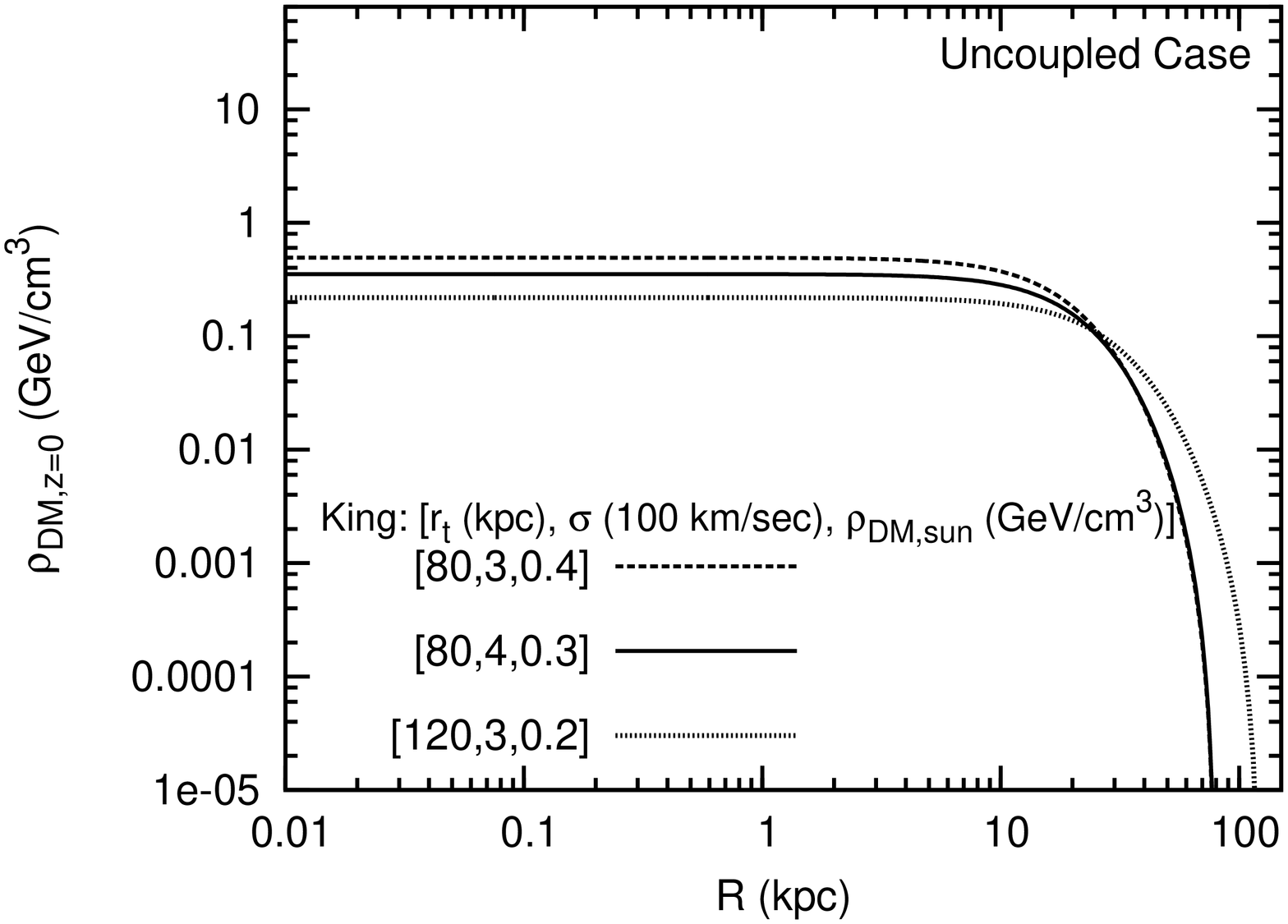,angle=270,width=3.5in}
&
\epsfig{file=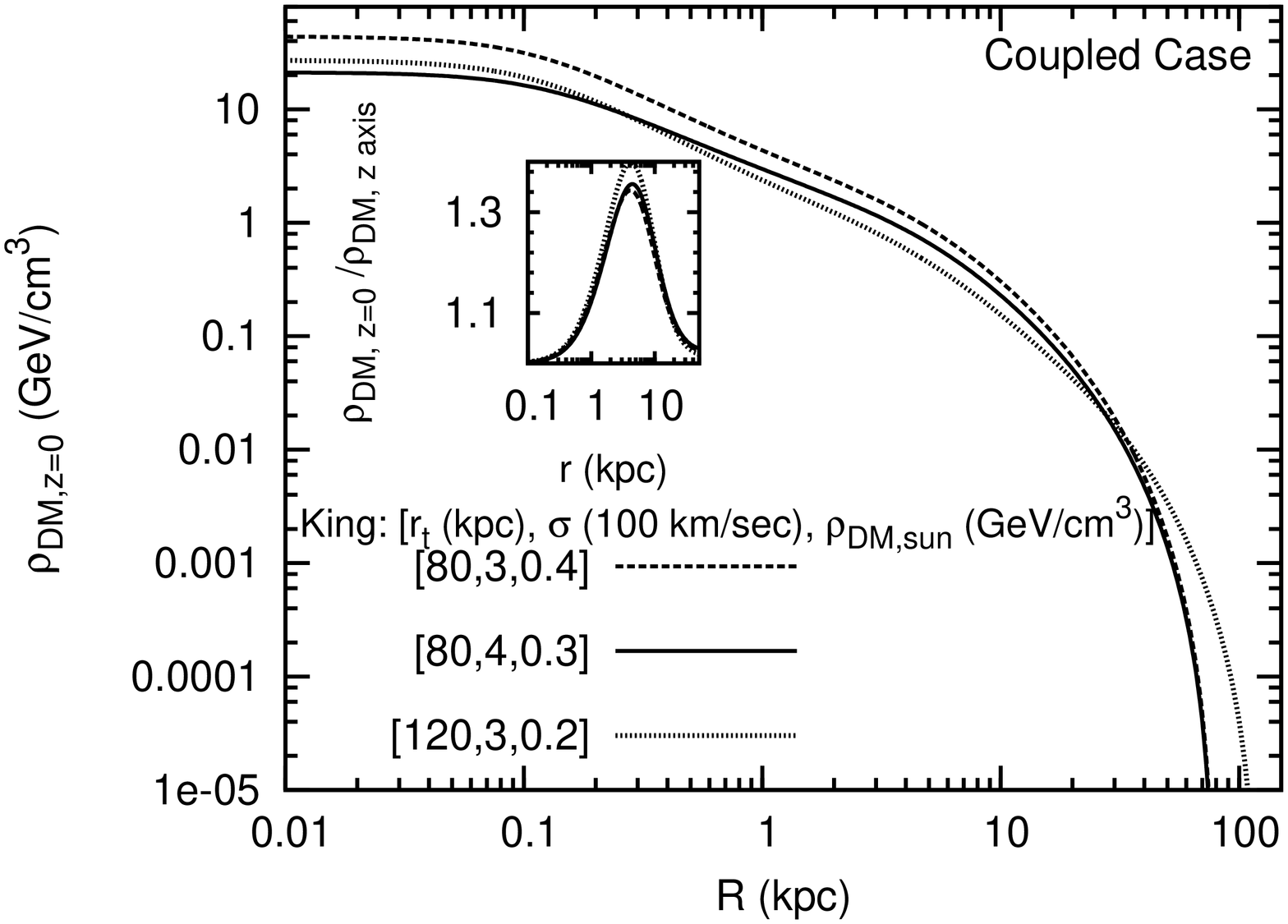,angle=270,width=3.5in}
\end{tabular}
\caption{The density profiles of the lowered (truncated) isothermal 
dark matter halo described by the King model DF, eq.~(\ref{eq:king_df}), 
(a) in absence (``uncoupled"), and (b) in presence (``coupled"), 
of the visible matter described by equations (\ref{eq:rho_sph}) and 
(\ref{eq:rho_disk}), for three different sets of values of the King 
model 
parameters ($r_t, \sigma, \rhodmsun$) that yield good fits 
(in the coupled case) to the rotation curve data of the Galaxy up to 
$\sim 60\kpc$. In (b) the density profiles refer to those on the 
equatorial plane ($z=0$) of the Galaxy. The inset in (b) shows the 
ratio of the density on the $z=0$ plane to 
that on the $z$ axis as a function of galactocentric distance in the 
coupled case, for the same chosen sets of the King model parameters as 
indicated.
}
\label{Fig:density_profiles}
\end{figure}

\noindent
The main effect of the gravitational influence of the visible matter on 
the DM halo, namely, increased central concentration 
and reduced core radius of the DM density profile, is illustrated  
in Figure \ref{Fig:density_profiles}. 
The left panel of Figure \ref{Fig:density_profiles} shows the density 
profiles of a `pure' spherically symmetric finite-size DM halo 
described by the King model DF (\ref{eq:king_df}) without including the 
gravitational influence of the visible matter on the DM (hereafter 
referred to as the ``uncoupled" case), whereas the right panel shows 
the DM density profiles (in the plane of the Galactic
disk ($z=0$)) under the additional gravitational 
influence of the visible matter (hereafter referred to as the 
``coupled" case). The curves in both panels are for three different 
sets of values of the King model parameters ($r_t, \sigma, \rhodmsun$) 
as indicated, that yield good fits (in the coupled case) to 
the rotation curve data of the Galaxy up to $\sim 60\kpc$ (see 
discussions below and Figure \ref{Fig:rot_curves}). The increased 
central density and the reduced core size of the density profiles in the 
coupled case relative to those in the uncoupled case are clearly seen. 

\paras 
Another direct effect of the gravitational influence of the visible 
matter on the DM is the enhancement of the DM density on 
the plane of the Galactic disk ($z=0$) relative to that off the disk. 
The inset in the right panel of Figure \ref{Fig:density_profiles} shows 
this effect clearly where it is seen that this disk enhancement of DM 
density can be as large as $\sim$ 30 -- 40 \%. The typical scale 
length of this ``dark matter disk" is a few kpc.   
\subsection{Rotation Curves
\label{subsec:rotcurves}}
\noindent
\begin{figure}[h]
\centering
\epsfig{file=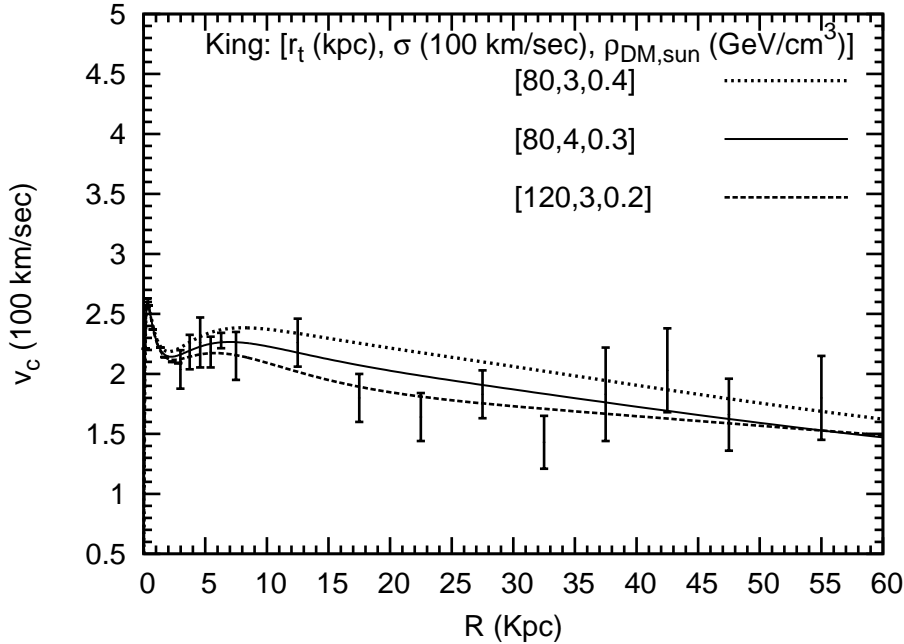,angle=270,width=5in}
\caption{Rotation curves for the Galaxy with its DM halo described by the 
King model DF, eq.~(\ref{eq:king_df}) and including the gravitational 
effect of 
the visible matter described by equations (\ref{eq:rho_sph}) and 
(\ref{eq:rho_disk}), for three different sets of values of the King 
model
parameters ($r_t\,, \sigma\,, \rhodmsun$) as indicated. The curves are
shown for three different values of $\rhodmsun$, and in each case, the curve
shown corresponds to the values of the other two parameters ($r_t$ and
$\sigma$) that yield best fit to the rotation curve data of the
Galaxy estimated in Ref.~\cite{rc_60kpc_Xue_etal_08}. The data with error 
bars are from Table 3 of Ref.~\cite{rc_60kpc_Xue_etal_08} (their data set 
$V_{\rm cir,  II}$ corresponding to their simulation II) for 
galactocentric distances $R\geq 7.5\kpc$, and from 
Ref.~\cite{Honma-Sofue_97} for $R < 7.5\kpc$. 
}
\label{Fig:rot_curves}
\end{figure}
Figure \ref{Fig:rot_curves} shows our theoretically calculated rotation 
curves for the Galaxy with its DM halo described by the King model DF, 
eq.~(\ref{eq:king_df}) and including the gravitational effect of the 
visible 
matter described by equations (\ref{eq:rho_sph}) and 
(\ref{eq:rho_disk}), 
for 
three different sets of values of the King model parameters ($r_t, 
\sigma, \rhodmsun$) as indicated. For comparison, the rotation curve 
data for $R$ up to 55 kpc given, for $R\geq 7.5\kpc$, in Table 3 of 
Ref.~\cite{rc_60kpc_Xue_etal_08} (their data set $V_{\rm cir,  II}$ 
corresponding to their simulation II) are shown together 
with the data from Ref.~\cite{Honma-Sofue_97} for $R < 7.5\kpc$. The 
theoretical rotation 
curves are shown for $\rhodmsun = 0.2\,, 0.3\,$ and $0.4\gev/\cm^3$, 
and in each case, the shown theoretical curve corresponds to the values 
of the other two parameters ($r_t$ and $\sigma$) that yield best fit 
(giving lowest $\chi^2$) to the above mentioned rotation curve data of 
the Galaxy.  

\begin{figure}[h]
\centering
\begin{tabular}{cc}
\epsfig{file=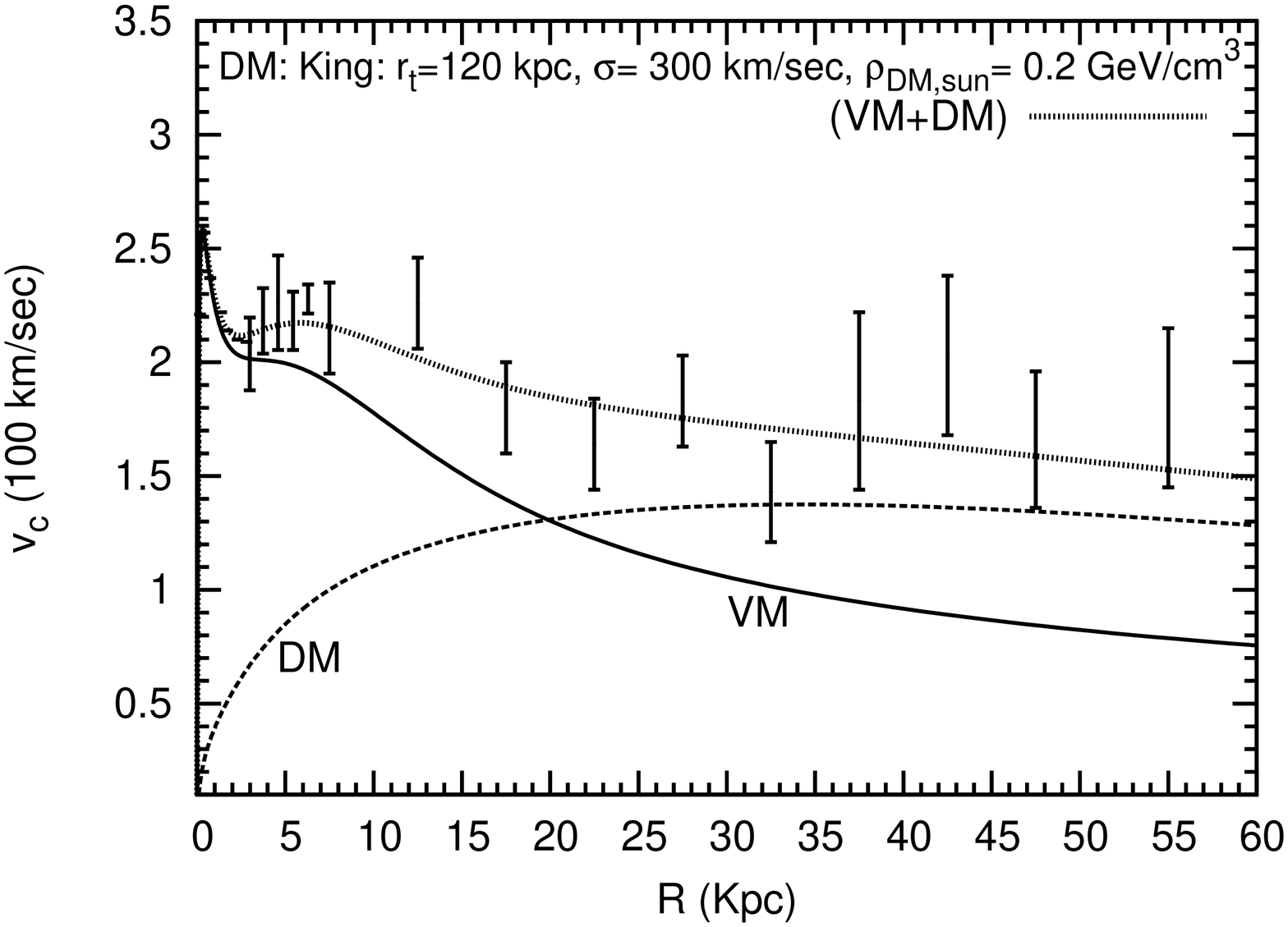,angle=270,width=3.5in}
&
\epsfig{file=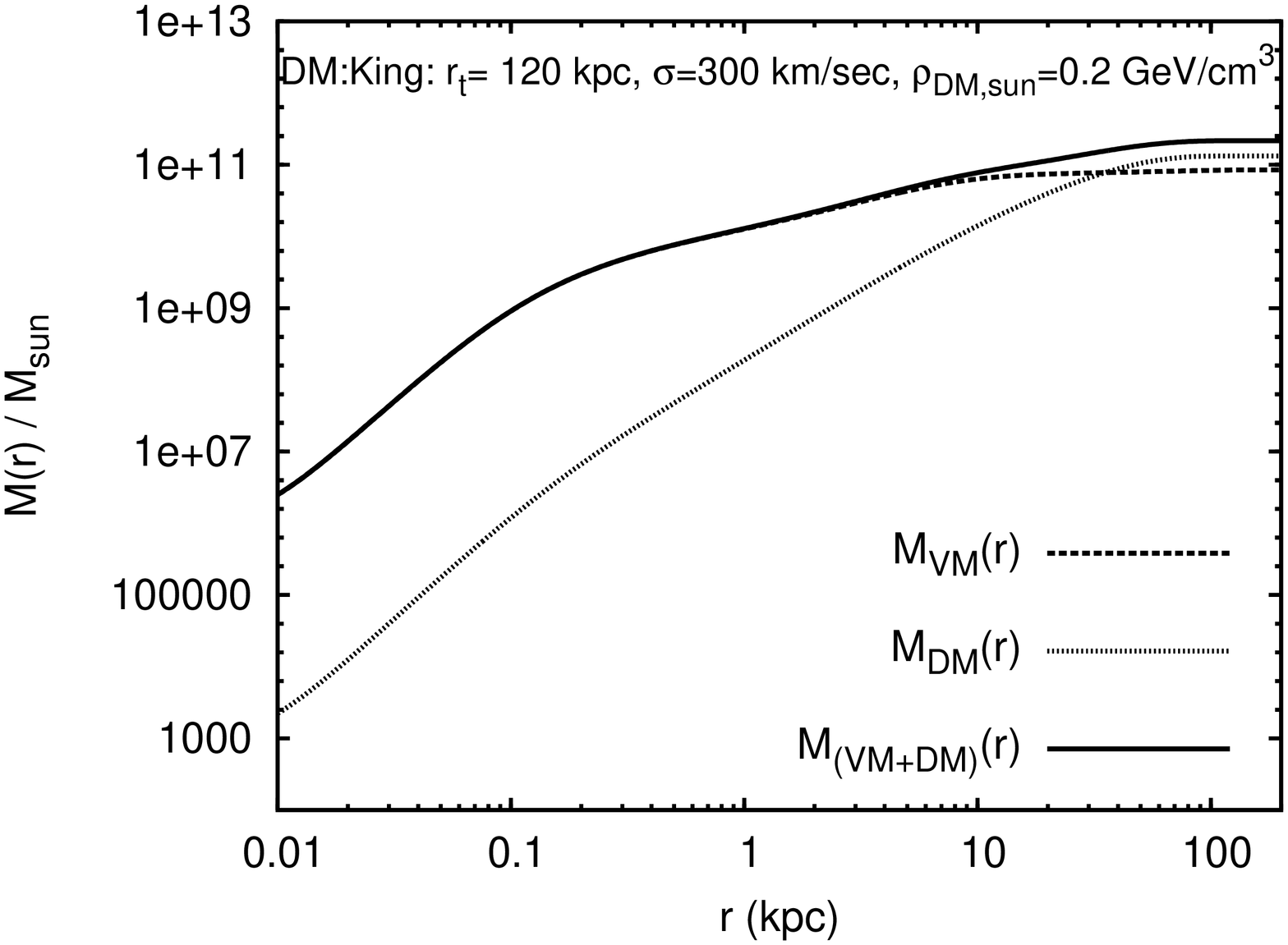,angle=270,width=3.5in}
\end{tabular}
\caption{The contributions of the visible matter (VM) and the dark 
matter (DM) components to (a) the total rotation curve (left panel) and 
(b) total mass (right panel) of the Galaxy for the best-fit King model 
DM halo parameters $\rhodmsun=0.2\gev/\cm^3$, $r_t=120\kpc\,$ and  
$\sigma\simeq 300\kmps$. The rotation curve data in the left panel are 
the same as in Figure \ref{Fig:rot_curves}. 
}
\label{Fig:rotcurve_and_mass_compo}
\end{figure}

\paras
For the range of King model density parameter considered ($0.2 \le 
\rhodmsun \le 0.4 \gev/\cm^3$), the ``global" best fit to the rotation 
curve data (giving globally lowest value of $\chi^2$) is obtained for 
the King model DM parameter values $\rhodmsun=0.2\gev/\cm^3$, 
$r_t\simeq 120\kpc$ and $\sigma\simeq 300\kmps$ (the dashed curve in 
Figure \ref{Fig:rot_curves})~\footnote{Note that, as already mentioned, 
the parameter $\sigma$ in the King model is not to be confused with the 
one-dimensional velocity dispersion of the DM particles.}. The separate 
contributions of the visible matter (VM) and the Dark Matter (DM) 
components to the total rotation curve and the total mass of the 
Galaxy for this ``best-fit" set of parameter values 
are shown in Figure \ref{Fig:rotcurve_and_mass_compo}.    

\paras 
It is of interest to note that the DM halo mass for the best-fit 
model obtained above, $M_{\rm DM}\sim 1.3\times10^{11}\msun$, is only 
marginally more than the total visible matter mass, $M_{\rm VM}\sim 
8.4\times 10^{10}\msun$, giving a total Galaxy mass of $\sim 
2.1\times10^{11}\msun$ for the adopted VM model. Actually, this is 
fairly independent of the exact values of the VM model parameters 
adopted as long as those parameters are such that the VM by itself gives 
the dominant contribution to the rotation curve data at small $R\ll 
\rsun$. The main reason for the relatively low DM halo mass is the 
declining nature of the rotation curve beyond the solar circle. Within 
the context of the 
truncated isothermal models of the DM halo described by the King 
models studied here, higher DM halo mass models ---  
which can be obtained by choosing larger values of one or 
more of the parameters $\rhodmsun$, $r_t$ or $\sigma$ than those of 
the best-fit model obtained above --- typically 
lead to rising or non-declining rotation curves beyond the solar circle, 
and are thus unable to explain the declining rotation curve data.     
\subsection{Maximum Speed and Speed Distribution of the particles 
\label{subsec:max_speed_and_speedDF}}
\begin{figure}[h]
\centering
\begin{tabular}{cc}
\epsfig{file=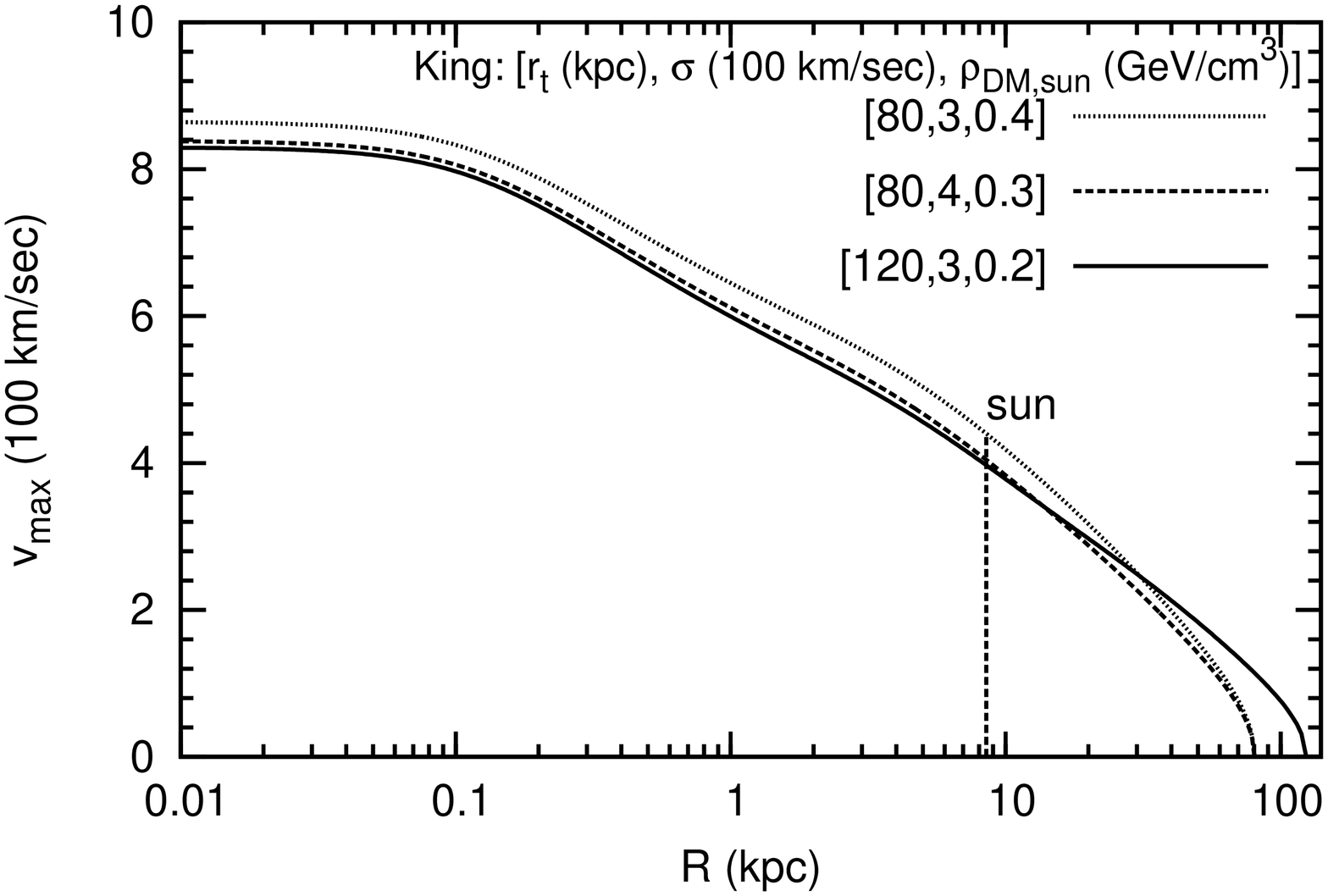,angle=270,width=3.5in}
&
\epsfig{file=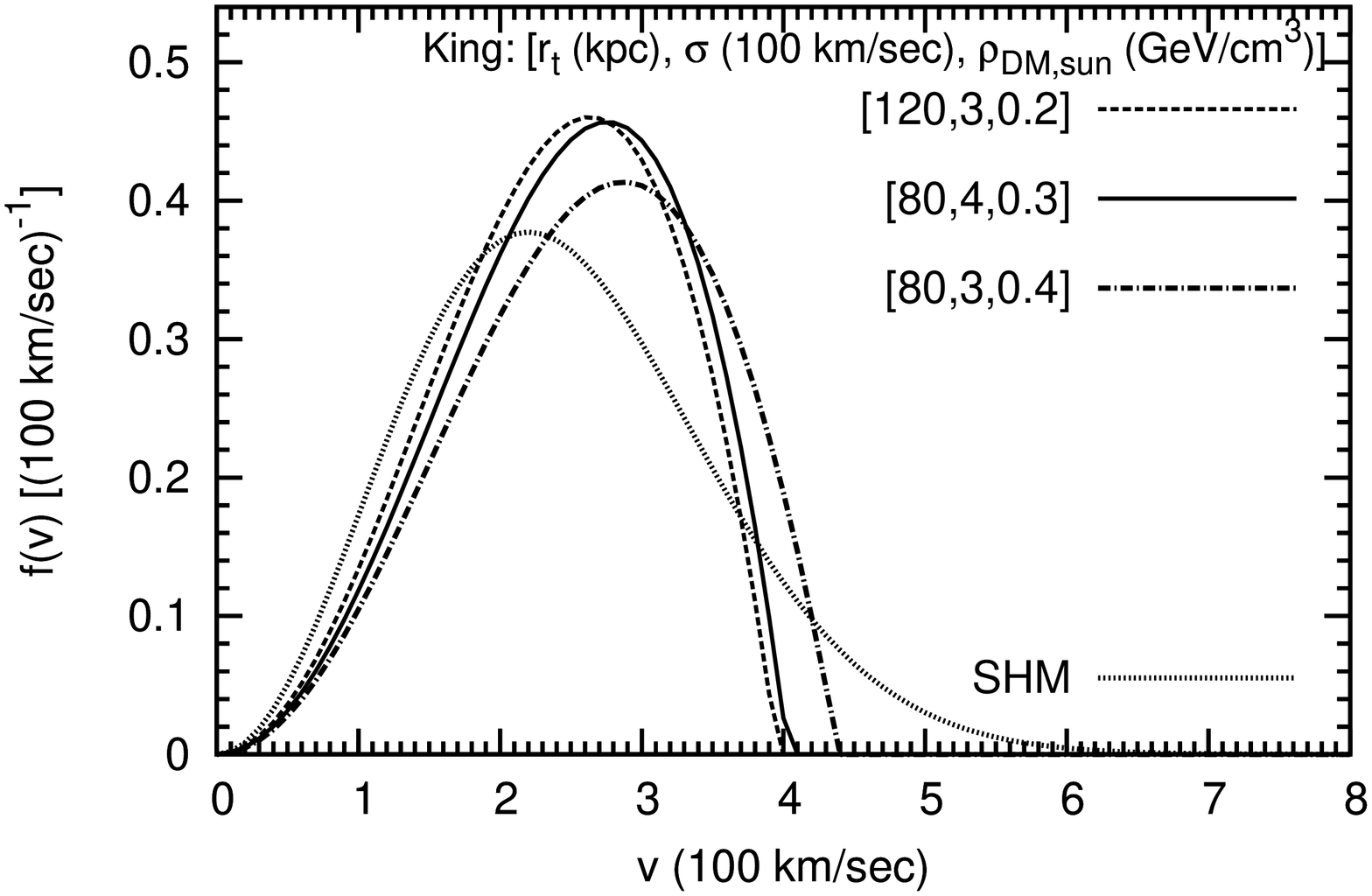,angle=270,width=3.5in}
\end{tabular}
\caption{Maximum speed $v_{\rm max} (\boldx)= \sqrt{2\Psi (\boldx)}$ in  
the Galactic 
equatorial plane as a function of Galactocentric distance $R$ (left panel) 
and the 
normalized speed ($v=|\boldv|$) distribution function of the DM particles, 
$f(v)\equiv \frac{4\pi v^2}{\rho(\boldx)} f(\boldx, \boldv)$ (right 
panel), at Sun's location $\left(R=\rsun,\, z=0\right)$, for the same three 
sets of King 
model parameters as in Figure \ref{Fig:rot_curves} as well as for the 
Maxwellian 
speed distribution in the case of Standard Halo Model (SHM).    
}
\label{Fig:maxspeed_and_DFatSun}
\end{figure}

\noindent
The self-consistently calculated maximum speeds of DM particles in 
the halo as a function of $R$ are shown in Figure 
\ref{Fig:maxspeed_and_DFatSun} (left panel) for the same three sets of 
King model 
halo parameters as in Figure \ref{Fig:rot_curves}. The corresponding 
normalized speed 
distribution functions of the DM particles (in the Galactic rest frame),  
$f(v)\equiv \frac{4\pi v^2}{\rho(\boldx)} f(\boldx, \boldv)$ 
(with $\int f(v) dv = 1$), 
at Sun's location $\left(R=\rsun,\, z=0\right)$, 
are also shown (right panel). For comparison, the Maxwellian speed 
distribution for the 
Standard Halo Model (SHM) at sun's location is also shown. Evidently, 
the speed distribution for the truncated ``isothermal" halo is 
significantly non-Maxwellian, especially at the high 
speed end of the distribution, not unlike the behavior found in recent 
numerical simulations~\cite{Ling_etal_simulations_09}. 

\paras
For the three King model parameter sets considered in the left panel of Figure 
\ref{Fig:maxspeed_and_DFatSun} the maximum speeds of the DM particles at 
the location of Sun are, from bottom to top, 396, 403 and 439 km/s, 
respectively. These values and the speed 
distribution functions shown in Figure \ref{Fig:maxspeed_and_DFatSun} will be 
used below in the calculation of the expected WIMP detection rates and 
analysis of the results of the direct detection experiments. We 
emphasize that the above values refer to the maximum speeds that a 
{\it dark matter particle} can have at the location of the Sun. Since 
stars in the Galaxy do not in general follow the same velocity 
distribution function as the dark matter particles, they can have 
speeds larger than the maximum speeds of the dark matter 
particles mentioned above. Stars with speeds larger than the escape 
speeds determined by the total mass and radius of the halo (see  
equation (\ref{eq:vescp_vmax_reln})), will, of course, 
eventually escape from the Galaxy. 

\section{Implications of the Self-consistent Truncated Isothermal Halo model 
for the Analysis of WIMP Direct Detection Experiments
\label{sec:dd_implications}}
\subsection{Direct Detection of WIMPs: The Basic Formalism
\label{subsec:Direct_Detection_Basics}}
\noindent 
For simplicity we restrict our attention in this paper to the case of 
elastic scattering of the WIMPs from nuclei.  
Consider a WIMP (hereafter denoted by $\chi$) of mass $\mchi$ 
scattering elastically from a nucleus of mass $M$. The nucleus 
receives 
a recoil energy, $E_R=q^2/(2M)=(\mu^2 u^2/M)(1-\cos\theta^*)$, 
where $\theta^*$ is 
the scattering angle in the centre of momentum frame, $q$ the 
momentum transferred to the recoiling nucleus, $\mu=\mchi  
M/(\mchi + M)$ is the WIMP-nucleus reduced mass and $u = |\boldu|$ is 
the speed of the WIMP relative to the nucleus assumed to be at rest on 
earth. The minimum WIMP speed that can produce a recoil energy $E_R$ of 
the nucleus is 
\beq
\umin (E_R)=\left(\frac{M E_R}{2\mu^2}\right)^{1/2}\,.
\label{eq:umin}
\eeq 
The differential event rate per unit detector 
mass (typically measured in counts/day/kg/keV) can be written as 
\cite{Jungman_etal_PhysRep_96,Lewin_Smith_96}
\beq
\frac{d \mathcalR}{dE_R}(E_R,t) = \frac{\sigma(q)}{2\mchi\mu^2}\rhochi 
\xi(E_R,t)\,,
\label{eq:recoil_rate_def}
\eeq
where $\rhochi\equiv\rhodmsun$ is the density of WIMPs in the solar 
neighborhood, $\sigma(q)$ is the 
WIMP-nucleus effective interaction cross section, and 
\beq
\xi(E_R,t)= \int_{u>\umin(E_R)}d^3\boldu 
\frac{\ftilde(\boldu,t)}{u}\,.  
\label{eq:xi_def}
\eeq
Here $\ftilde(\boldu,t)$ is the time-dependent velocity 
distribution of the WIMPs in the solar neighborhood {\it relative to the 
detector at rest on earth}, and $\umin$ is defined in equation 
(\ref{eq:umin}). Recall that  
$f_{\rm K}(\boldx,\boldv)$ defined in equation (\ref{eq:king_df}) refers 
to the rest frame of the Galaxy, and 
has no explicit dependence on time. The time dependence of $\ftilde$ arises 
from the motion of the earth with respect to the Galactic rest frame. 
The two distribution functions are simply related as 
\beq
\ftilde(\boldu,t) = \frac{1}{\rhochi}f_{\rm 
K}\left(\boldx=\boldx_\odot,\boldv=\boldu+\boldvE(t)\right)\,,
\label{eq:ftilde_f_relation}
\eeq     
where $\boldx_\odot$ represents the sun's position ($R=8.5\kpc, z=0$) 
and $\boldvE(t)$ is the earth's velocity vector {\it in 
the Galaxy's rest frame}. Note that the DF $f_{\rm K}$ vanishes for 
speed $v\ge\vmax$ defined in equation (\ref{eq:vmax}). 

\paras 
Using equation (\ref{eq:ftilde_f_relation}) we can explicitly write 
equation (\ref{eq:xi_def}) as 
\beq
\xi(E_R,t)= \frac{2\pi}{\rhodmsun} \int_{-1}^1 d(\cos\theta)\,\, 
\Theta\left(\umax - \umin\right)
\int_{\umin(E_R)}^{\umax(\cos\theta)} u  f_{\rm
K}\left(\boldx=\boldx_\odot,\boldv=\boldu+\boldvE(t)\right) du\,,
\label{eq:xi_def2}
\eeq
where $\umax$ is the positive root of the equation 
\beq
\vmax^2 = \umax^2 + \vE^2 + 2\umax \vE\cos\theta\,, 
\label{eq:umax_def}
\eeq
with $\vmax$ given by equation (\ref{eq:vmax}) evaluated at 
$\boldx=\boldx_\odot$, and $\vE = |\boldvE|$ given by 
\cite{Petriello-Zurek_08} 
\beq
\vE (t) = \vsun + \vEorb \cos\gamma \cos\left[\omega(t-t_0)\right]\,,
\label{eq:vE_def}
\eeq 
where $\vsun \approx \vcsun + 12\kmps\,$ is the magnitude of the 
dominant component of sun's 
total velocity vector (which includes its circular velocity around the 
Galaxy of magnitude $\vcsun\approx220\kmps$ and its 
``peculiar" velocity of magnitude $\approx 12\kmps$) in the 
Galactic rest frame, $\vEorb\approx 30\kmps$ is the average orbital 
speed of the earth around the sun, $\cos\gamma\approx 0.51$, 
$\omega=2\pi/365$ and $t_0=152.5$ (corresponding to 2nd June when the 
earth's speed with respect to the Galactic rest frame is maximum), $t$ 
being counted in days. It is the periodic variation of $\vE$, equation 
(\ref{eq:vE_def}), that gives rise to the annual modulation signal 
claimed 
to have been detected by the DAMA collaboration.  

\paras 
For detectors composed of more than one kind of elements, the 
total differential event (i.e., nuclear recoil) rate is  
\beq
\left(\frac{d\mathcalR}{dE_R}\right)_{\rm tot} = \sum_i f_i 
\left(\frac{d\mathcalR}{dE_R}\right)_i\,,
\label{eq:recoil_rate_total} 
\eeq
where $f_i$ is the mass fraction of, and 
$\left(\frac{d\mathcalR}{dE_R}\right)_i$ 
the differential event rate (eq.~(\ref{eq:recoil_rate_def})) for, the 
nuclear species $i$, with corresponding nuclear mass $M_i$, 
WIMP-nucleus cross section $\sigma_i(q)$ and 
WIMP-nucleus reduced mass $\mu_i$. 

\paras
The number of nuclear recoil events in a recoil energy range between 
$E_R^1$ and $E_R^2$ is 
\beq
N_R (E_R^1, E_R^2) = \sum_i \int_{E_R^1}^{E_R^2}\,dE_R\, 
\left(\frac{d\mathcalR}{dE_R}\right)_i\, \mathcalK_i (E_R)\,, 
\label{eq:events_ER1_ER2}
\eeq
where $\mathcalK_i = \mathcalM_i T \epsilon (E_R)$ is the total exposure 
of 
the detector, $\mathcalM_i$ being the total mass of the species $i$ in 
the detector, $T$ the total exposure time, and $\epsilon (E_R)$ the 
energy-dependent detector efficiency. 

\paras
The WIMP-nucleus total effective scattering cross section, $\sigma(q)$, 
can be written as a sum of two contributions arising from 
spin-independent (SI) and spin-dependent (SD) effective couplings of the 
WIMP to the detector 
nucleus~\cite{Jungman_etal_PhysRep_96,Lewin_Smith_96,Savage_etal_08}~: 
\beq
\sigma(q) = \sigma_{\rm SI}(q) + \sigma_{\rm SD}(q)\,.
\eeq

\paras
{\it Spin-independent} (SI) (or {\it Coherent}) Scattering~:~ In this 
case, it is assumed that the WIMP interacts coherently with the nucleus 
as a whole. The cross section is 
generally written as~\cite{Jungman_etal_PhysRep_96} 
\beq
\sigma (q) = \sigma_0 |F(q)|^2\,
\eeq 
where $\sigma_0$ is the zero-momentum WIMP-nucleus scattering cross 
section, and $F(q)$ is a momentum dependent form factor that 
arises from the finite size of the nucleus. The form factor $F(q)$ is 
normalized to $F(0) = 1$. The coherent nature of the interaction is 
implemented under the usual assumption that $\sigma_0$ scales with the 
square of the atomic mass number $A$ (number of protons plus neutrons) 
of the nucleus. For purely scalar interactions, with equal 
WIMP couplings to protons and neutrons, one can write $\sigma_0$ in 
terms of the WIMP-proton (or WIMP-neutron) effective cross section 
$\sigma_{\chi\, p} = \sigma_{\chi\, n}$ as  
\beq
\sigma_0^{\rm SI} = \sigma_{\chi\, p}^{\rm SI} 
\left(\frac{\mu}{\mu_{\chi\, p}}\right)^2 A^2\,,
\label{eq:sigma_0_SI}
\eeq 
where $\mu_{\chi\, p}$ is the WIMP - proton reduced mass. 

\paras
We shall use the conventional Helm form of the nuclear form 
factor~\cite{Jungman_etal_PhysRep_96,Lewin_Smith_96,Savage_etal_08}~: 
\beq
F(q) = 3 e^{-q^2 s^2/2\hbar^2} \, \frac{\sin(qr_0/\hbar)- 
(qr_0/\hbar)\cos(qr_0/\hbar)}{(qr_0/\hbar)^3}\,,
\label{eq:FormFactor_SI}
\eeq
with $s=0.9\,$fm, the nuclear skin thickness and $r_0=1.14 
A^{1/3}\,$ fm, the effective nuclear radius. 
 
\paras
{\it Spin-dependent} (SD) Scattering~:~ In this case the WIMP couples to 
the total spin $J$ of the nucleus which has contributions from the 
spins of the individual protons and neutrons within the nucleus. 
Following Ref.~\cite{Savage_etal_08} we use 
the effective WIMP-nucleus SD cross section generically 
written in the form~\cite{Engel_PLB_91}
\beq
\sigma_{\rm SD}(q) = \frac{32 \mu^2 G_F^2}{2 J + 1} 
\left[a_p^2 S_{pp}(q) + a_p a_n S_{pn}(q)
+ a_n^2 S_{nn}(q) \right]\,,
\label{eq:sigma_SD}
\eeq 
where $a_p$ and $a_n$ are respectively the axial four-fermion 
WIMP-proton and WIMP-neutron couplings in units of $2\sqrt{2}G_F$  
\cite{Gondolo_hepph_9605290,Tovey_etal_PLB_00,Gondolo_etal_JCAP_04}. The 
nuclear structure functions $S_{pp}(q)$, $S_{nn}(q)$ and
$S_{pn}(q)$ are taken, for Ge from Ref.~\cite{Dimitrov_etal_PRD_95}, 
for Si from Ref.~\cite{Ressell_PRD_93}, for Al from 
Ref.~\cite{Engel_etal_PRC_95}, and for Na, I and Xe from 
Ref.~\cite{Ressel&Dean_PRC_97}.  

\paras
In this paper, following the standard practice, we shall study the 
situations when either $a_n=0$ or $a_p=0\,$; the relevant effective 
cross 
section in the two cases will be denoted by $\sigma_{\chi \,p}^{\rm 
SD}$ and $\sigma_{\chi \,n}^{\rm SD}$, respectively. In general both 
$a_n$ and $a_p$ can be non-zero, but we will not consider this case 
here.    

\subsection{Results of Direct Detection Experiments and Analysis 
Techniques
\label{subsec:Expts}}
\noindent 
To illustrate the general implications of our halo model for the WIMP 
mass and cross section as implied by the results of direct-detection 
experiments, we consider in this paper the results of three of the 
``null" experiments (i.e., those without any definite claim of  
detection so far), namely, 
CDMS-II~\cite{cdms-II_Si_PRL_06,cdms-II_Ge_PRL_10}, 
CRESST~\cite{cresst} and XENON10~\cite{xenon10a,xenon10b}, in 
addition to those of the DAMA collaboration~\cite{dama_libra_08} which 
has claimed a positive signal based on the claimed detection of the 
annual modulation of the recoil event rate. 
As we shall see, results of these three null experiments together 
provide useful information regarding the question 
of compatibility of the positive results of DAMA with the negative 
results of other experiments over a wide range of the WIMP mass $\mchi$ 
from $\sim$ 1 GeV to 100 GeV. Below, we first describe the DAMA results 
and the analysis procedure we follow in order to derive the constraints 
on WIMP mass and interaction cross section imposed by the results, and 
then discuss the same for the null experiments. 

\paras 
{\it DAMA/NaI and DAMA/LIBRA}~: \\
The DAMA collaboration~\cite{dama_NaI,dama_libra_08} has claimed 
detecting a non-zero annual 
modulation signal at a confidence level (C.L) of $8.2\sigma$ in their 
number of detected nuclear recoil events over a period of seven  
(for the DAMA/NaI experiment~\cite{dama_NaI}), plus four (for the 
DAMA/LIBRA experiment~\cite{dama_libra_08}),  
annual cycles, with a total exposure of 0.82 ton-year\footnote{While 
this paper was being prepared, the DAMA/LIBRA collaboration  
released their latest results with an additional exposure of 0.34 
ton-year corresponding to two additional annual 
cycles~\cite{dama_libra_10}. With this, the  
statistical significance of the annual modulation signal is now 
$8.9\sigma$ for the cumulative exposure. We do not 
include these new results in this paper.}. The DAMA collaboration 
attributes this annual modulation to the periodic variation of $\vE$, 
equation (\ref{eq:vE_def}), which results in a periodic variation of the 
recoil event rate (see discussions in the previous subsection). 

\paras
Given the time dependence of $\vE$, equation (\ref{eq:vE_def}), the time 
dependence of the differential event rate 
(\ref{eq:recoil_rate_def}) due to motion of the earth around the sun can 
be 
approximately written as~\cite{DFS_86,FFG_88,Savage_etal_08}
\beq
\frac{d \mathcalR}{dE_R}(E_R,t)\approx 
S_0(E_R)+S_m(E_R) \cos \omega (t-t_0)\,,
\label{eq:recoil_rate_modu_def}
\eeq
where $S_0$ is the average recoil rate over a year and $S_m$ is  
the ``modulation amplitude" defined as 
\beq
S_m(E_R)=\half\left[\frac{d\mathcalR}{dE_R}\left(E_R,\, {\rm June 
2}\right) - \frac{d\mathcalR}{dE_R}\left(E_R,\, {\rm Dec 2}\right) 
\right]\,.
\label{eq:S_m_def}
\eeq
A nonzero value of $S_m$ is taken to be a signal for WIMP-induced 
recoils.   

\paras
To compare with DAMA data given in specific recoil energy bins, the 
average value of $S_m$ over a given energy range is calculated as:

\beq
S_m^{E_1-E_2}=\frac{1}{E_2-E_1}\int_{E_1}^{E_2}{S_m(E_R)dE_R}\,. 
\label{eq:S_m_bin_average}
\eeq

\paras
A scintillation detector such as that used in the DAMA experiment 
directly detects only the part of the recoil nucleus energy that 
goes into electromagnetic channel and produces the observed 
scintillation. The actual energy of the recoil nucleus $E_R$ is related 
to the detectable energy from the scintillation light yield, 
$E_D$ --- often called the ``electron-equivalent (ee) energy" and 
denoted by ``keVee" with energy in units of keV ---   
through the relation $E_D = Q E_R$, where the ``quenching factor" $Q 
(<1)$ depends on the nuclear 
material composing the scintillation detector. For the DAMA detector 
(composed of NaI crystal), $Q_{\rm Na}\approx 0.3$ and $Q_{\rm I}\approx 
0.09$. At the same time, as first pointed out in 
Ref.~\cite{Drobyshevski_channeling_07} and studied in detail in the 
context of the DAMA detector in 
Ref.~\cite{Bernabei_etal_DAMA_channeling_08}, for certain energies 
and incidence angles of the particle (for example, along the crystal 
axis), the incident particle transfers energy only to the electrons 
(rather than to the nuclei) of the scintillator material. In such a 
case, called ``channeling", one has $Q\approx 1$. We shall use the 
following simple parametrizations, suggested in 
Ref.~\cite{Foot_channeling_frac_08}, of 
the fractions of ``channeled" events for recoiling Sodium and Iodine 
nuclei, based on simulation results given 
in Ref.~\cite{Bernabei_etal_DAMA_channeling_08} for the DAMA experiment:  
\beq
f_{\rm Na} \simeq \frac{1}{1+1.14E_R(\kev)}\,, \,\,\,\,\,\,\, 
f_{\rm I} \simeq \frac{1}{1+0.75E_R(\kev)}\,.
\label{eq:channel_frac}
\eeq

\paras
Note that the use 
of keVee is only a bookkeeping device to distinguish the actually 
measured energy by the detector from the true recoil energy, the  
actual energy {\it unit} still being keV. For 
detectors for which the concept of channeling does not exist (for 
example, CDMS or XENON), the quenching factor can be included in the 
energy calibration and final data quoted directly in terms of the recoil 
energy. 

\paras
For the analysis of the DAMA results, in this paper we consider for 
simplicity only the 2-bin data set given by the DAMA 
collaboration~\cite{dama_libra_08}, namely, the low-energy bin 2 -- 6 
keVee within which a non-zero modulation amplitude is measured, and the 
high energy bin 6 -- 14 keVee in which the modulation amplitude is 
consistent with zero. Table I gives the modulation 
amplitudes ($S_m$) measured by DAMA in these two energy bins. 

\begin{table}[h]
\begin{center}
\begin{tabular}{|c|c|}\hline
Energy (keVee) & Modulation Amplitude ($S_m$) (counts/day/kg/keVee) \\ 
\hline
2 -- 6 & 0.0131 $\pm$ 0.0016 \\
6 -- 14 & 0.0009 $\pm$ 0.0011\\
\hline
\end{tabular}
\caption{DAMA modulation amplitude data~\cite{dama_libra_08}}
\end{center}
\end{table}

\paras 
The efficiency of the DAMA detector is taken to be unity. To 
compare with the actual experimental data we take into account the 
finite energy resolution of the detector whereby the actually measured 
energies are taken to be normally distributed about the true detectable 
energy $E_D$ (i.e., the energy that would be measured if the 
detector had 100\% energy resolution) with a standard 
deviation~\cite{dama_energy_resol_NIM_08}
$\sigma(E_D)=(0.448\kev)\sqrt{E_D/\kev} + 0.0091 E_D\,$. Thus, the 
expected modulation amplitude is obtained by convolving 
equation (\ref{eq:S_m_def}) (after changing the integration variable to
$E_D=QE_R$) with a normalized Gaussian with 
the above standard deviation. The expected modulation amplitude over 
a given interval of measured energy between $E_{D1}$ and $E_{D2}$ is 
then calculated from equation (\ref{eq:S_m_bin_average}) after the above 
convolution. 

\paras
With the theoretically expected modulation amplitude in the $k$-th 
energy bin, $S_{m,k}^{\rm th}$, calculated as described above, we 
perform, following the simple  
analysis procedure of Ref.~\cite{Petriello-Zurek_08}, a $\chi^2$ fit to 
the experimental modulation amplitude data given in Table~I with 
\beq
\chi^2\equiv\sum_k\left(\frac{S_{m,k} - S_{m,k}^{\rm 
th}}{\sigma_k}\right)^2\,,
\label{eq:chisq_def}
\eeq
where $S_{m,k}$ is the experimentally measured modulation amplitude in 
the $k$-th energy bin and $\sigma_k$ the corresponding error in the 
experimental value given in Table~I. In this simple analysis 
procedure, there are only two free parameters in the 
problem, namely, the WIMP mass $\mchi$ and the relevant WIMP-nucleon 
cross section $\sigma_\chi$. The latter is $\sigma_{\chi\, p}^{\rm SI}$ 
in the SI case (see eq.~(\ref{eq:sigma_0_SI})), and 
either $\sigma_{\chi\, p}^{\rm SD}$ or $\sigma_{\chi\, n}^{\rm SD}$  in 
the SD case (see eq.~(\ref{eq:sigma_SD}) and the discussions 
following it). For a given value of $\mchi$ we find $\chi^2_{\rm 
min}$, the minimum value of 
$\chi^2$, by scanning over the values of the relevant cross section 
$\sigma_\chi$. The 90\% C.L. allowed region of the relevant cross 
section, for the given value of $\mchi$, is then found by accepting 
those values of $\sigma_\chi$ for which $\chi^2 - \chi^2_{\rm
min} \leq 2.71$, provided $\chi^2_{\rm min} < 2$. More rigorous 
multi-parameter analysis procedures described, for instance, in 
Ref.~\cite{Savage_etal_08}, may alter the constraints derived here at 
some quantitative level. However, we believe the simple analysis 
procedure adopted here is sufficient for our main purpose of 
illustrating, in the context of our self-consistent model of the 
dark halo of the Galaxy, the general nature of the constraints 
in the ($\mchi,\, \sigma_\chi$) parameter space implied by the 
experimental results. The resulting constraints imposed by the DAMA 
results are described in the next subsection together with those imposed 
by the null results of the other experiments. 

\paras
{\it Experiments giving null results}~:\\
Table II summarizes the relevant features of the ``null" experiments we 
consider in this analysis. Details of each experiment can be found in 
the cited References. 

\begin{table}[h]
\begin{center}
\begin{tabular}{|c|c|c|c|}\hline
Experiment&Target&Effective exposure (kg-days)&Threshold (keV) \\ \hline
CDMS-II~\cite{cdms-II_Ge_PRL_10} & Ge & 304.5 (SI), 23.5 (SD) & 10\\
CDMS-II~\cite{cdms-II_Si_PRL_06} & Si & 12.1 (SI), 0.57 (SD) & 7\\
CRESST-I~\cite{cresst} & ${\rm Al}_2 {\rm O}_3$ & 1.51 (SI), 0.53 (SD) 
& 0.6\\
XENON10~\cite{xenon10a,xenon10b} & Xe & 136 (SI), 64.7 (SD) & 
6.1\\\hline
\end{tabular}
\caption{Relevant features of the experiments with null results 
considered in this paper. 
}
\end{center}
\end{table}

\paras 
For the purpose of analysis, the null experiments can be divided 
into two classes: (a) Those which report no events at all (after all the 
relevant cuts are employed), and (b) those which report events (after 
analysis cuts) but ascribe them to background. In both cases, we follow 
the simple analysis procedure outlined in 
Ref.~\cite{Petriello-Zurek_08}. For the former case (a), 
we derive 90\% C.L.~upper limit on the relevant WIMP cross section 
$\sigma_\chi$ for a given WIMP mass $\mchi$ by using Poisson statistics 
and 
demanding that the theoretically predicted number of events, $N$, over 
the entire energy range of interest be such as to allow a Poisson 
probability of observing zero events as bad as 10\% but not worse. This 
corresponds to $N(\mchi,\sigma_\chi)=2.3$. In the case (b), a simplified 
version of Yellin's~\cite{Yellin_PRD_02} optimum interval method, as 
suggested in Ref.~\cite{Petriello-Zurek_08}, is used, whereby we again  
calculate, for a given mass $\mchi$, and for all contiguous 
combination of energy bins, the 90\% C.L.~upper limit allowed 
theoretically expected number of events $N(\mchi,\sigma_\chi)$ 
corresponding 
to the observed number of events $n$ as dictated by Poisson statistics, 
i.e., by the formula $\sum_{j=0}^{j=n}\frac{N^j}{j!}\exp(-N) = 0.1$, 
and choose the most stringent (i.e., the lowest) upper limit value of 
$\sigma_\chi$ so obtained.      

\paras
CDMS-II (Ge)~\cite{cdms-II_Ge_PRL_10}~:~\\
This is currently the experiment with the largest effective exposure 
amongst all currently running experiments. Two events were observed, 
after applying all cuts, in the signal region 
at recoil energies 12.3 keV and 15.5 keV in the 10 -- 100 keV window. The 
effective quenching factor is already included in the energy 
calibration. Since the estimated probability of observing 2 or more 
background events in this window is $\sim$ 23\%, 
these events were ascribed to background. We obtain upper limits on the 
relevant WIMP cross section as a function of WIMP mass using the 
simplified version of Yellin's~\cite{Yellin_PRD_02} optimum interval 
method described above. In this analysis we have used a minimum bin 
width of 0.4 keV. The total effective exposure used in 
this calculation includes the spectrum-averaged equivalent exposure of 
$\approx$ 194.1 kg-days (for a WIMP mass of $\sim 60\gev/c^2$) 
reported in Ref.~\cite{cdms-II_Ge_PRL_10} combined with 
that of CDMS-II collaboration's previous published 
paper~\cite{cdms-II_Ge_PRL_09}, $\sim$ 
110.4 kg-days (after reduction of the figure quoted in 
Ref.~\cite{cdms-II_Ge_PRL_09} by a factor of $\sim$ 9\% due to improved 
estimate of their detector mass as prescribed in 
Ref.~\cite{cdms-II_Ge_PRL_10}). The above figures include the 
approximately 
constant (in energy) detector efficiency of $\sim$ 30\% already folded 
in.  

\paras
For spin-dependent interaction, only the $^{73}{\rm Ge}$ isotope, 
whose natural abundance is $\approx$ 7.73\%, is 
sensitive to the WIMPs. The effective exposure for SD interaction is 
correspondingly reduced and is significantly less than that for the SI 
interaction case. 

\paras
CDMS-II (Si)~\cite{cdms-II_Si_PRL_06}~:~\\
This experiment employs a high purity silicon crystal as the detector 
material. No events passed all the data analysis cuts. We 
derive 90\% C.L. upper limit on the relevant cross section as a 
function of the WIMP mass by using Poisson statistics discussed above, 
i.e., by demanding that, for a given WIMP mass $\mchi$, the upper limit 
value of cross section yield 2.3 events. 

\paras
For SD interactions, only the $^{29}{\rm Si}$ isotope, whose natural 
abundance is 4.68\%, is effective.  

\paras
CRESST-I~\cite{cresst}~:~\\
The Phase I of the CRESST experiment employed sapphire (${\rm Al}_2{\rm 
O}_3$) detectors with an exposure of 1.51 kg-days over an energy range 
of 0.6-20 keV. Measured energies are calibrated to recoil energy ($Q=1$).   
All the observed events were ascribed to background. We take into 
account the energy resolution of the detector described by a Gaussian 
distribution of the measured energies around the true value with a 
standard deviation $\sigma(E_R) = \sqrt{(0.220\kev)^2+(0.017E_R)^2)}$ 
~\cite{cresst}\footnote{Ref.~\cite{cresst} actually quotes their energy 
resolution in terms of the FWHM, $\Delta {E_R}^{\rm FWHM}$, which is 
related to the standard deviation $\sigma$ by $\Delta {E_R}^{\rm 
FWHM}\approx 2.354\sigma$.}. Upper limits were obtained
similar to CDMS-II(Ge) analysis with a smallest allowed energy-interval 
width of 1.2 keV~\cite{cresst}. 

\paras
For SD case only the Aluminum is sensitive since Oxygen has J=0.

\paras
{\it XENON10}~\cite{xenon10a,xenon10b}~:~\\
XENON10 is a dual phase (liquid and gas) xenon time projection chamber 
and uses the ratio of ionization to scintillation yield for 
discriminating between the dominant electron-recoil background and the 
looked for nuclear-recoil WIMP signal over a nuclear-recoil energy range 
of $\sim$ 6.1--36.5 keV ($Q=1$) with the relevant energy 
calibration parameter ${\mathcal L}_{\rm 
eff}=0.14$~\cite{Savage_etal_08}. We use an energy resolution of 
$\sigma(E_R)=(0.579\kev)\sqrt{E_R/\kev} + 0.021E_R$ as suggested in 
Ref.~\cite{Savage_etal_08}. A total of 10 
events (without background subtraction) were recorded in the WIMP signal 
region for a total effective exposure of about 136 kg-days after 
analysis cuts. As in the case of the CDMS-II Ge analysis described 
above, we again obtain upper limits on the relevant WIMP cross section 
as a function of WIMP mass using the simplified version of 
Yellin's optimum interval method described above.

\paras
For spin-dependent WIMP interaction, only the isotopes $^{129}{\rm Xe}$ 
(spin-1/2) with a natural abundance of $\sim$ 26.4\% and $^{131}{\rm 
Xe}$ (spin-3/2) with a natural abundance of $\sim$ 21.2\% are effective. 

\subsection{Implications for the WIMP parameters~:~Exclusion plots and 
compatability of null results with DAMA signal
\label{subsec:exclusion_plots}}
\begin{figure}[h]
\begin{center}
\epsfig{file=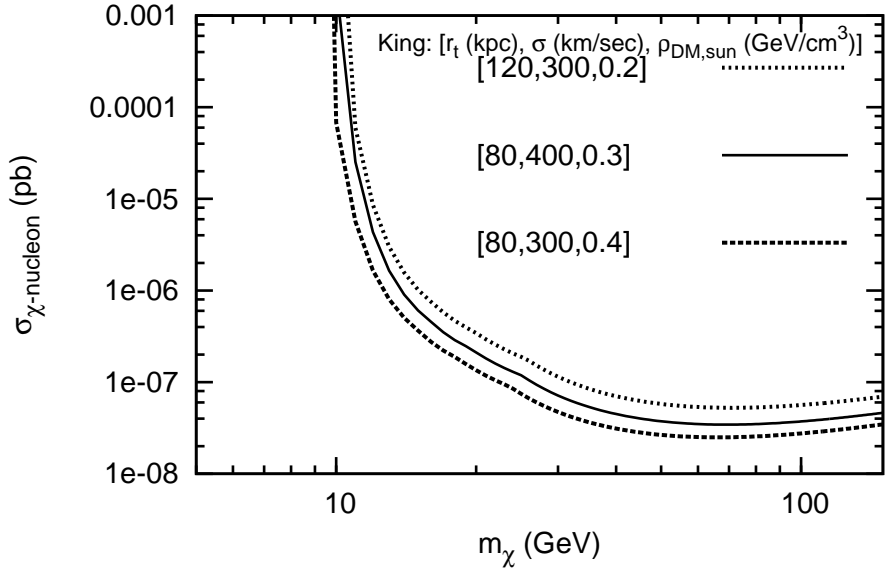,angle=270,width=5in}
\caption{90\% C.L. upper limits on the WIMP-nucleon spin-independent 
cross section as a function of WIMP mass as implied by the latest CDMS 
results~\cite{cdms-II_Ge_PRL_10}, for our self-consistent truncated 
isothermal (King) model of Milky Way's DM halo described by the same 
three sets of King model parameters as in Figure \ref{Fig:rot_curves}. 
The WIMP speed distribution function and the maximum WIMP speeds for 
these King model parameters used in deriving the above upper 
limits are as shown in Figure 
\ref{Fig:maxspeed_and_DFatSun}. 
}
\label{Fig:exclusionplot_CDMS}
\end{center}
\end{figure}

\begin{figure}[h]
\begin{center}
\epsfig{file=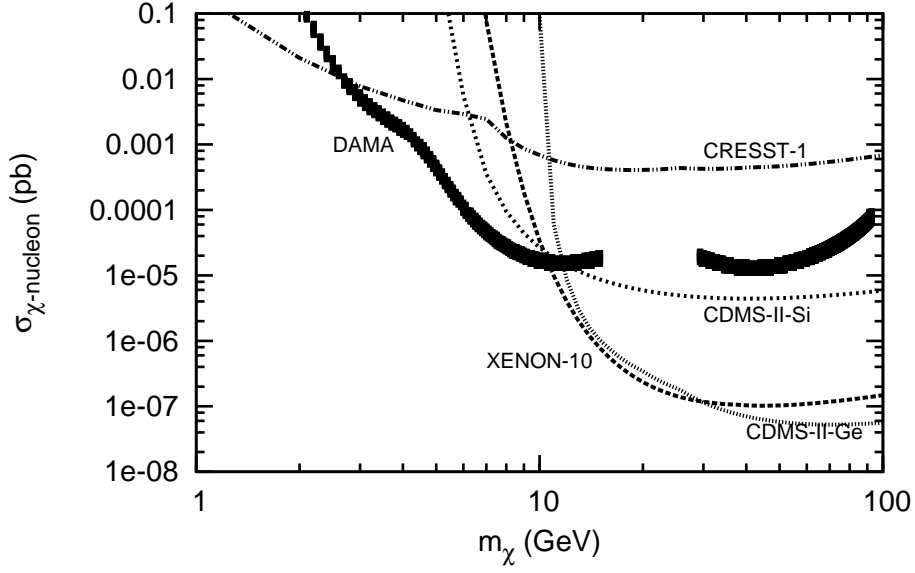,angle=270,width=5in}
\caption{90\% C.L. allowed region in the WIMP mass vs.~WIMP-nucleon 
spin-independent cross section plane implied by the DAMA modulation 
signal, for our best-fit halo model with parameters $r_t=120\kpc$, 
$\sigma=300\kmps$ and $\rhodmsun=0.2\gev/\cm^3$. Also shown are 90\% 
C.L. upper limits on cross 
section as a function of WIMP mass as implied by the null results of other 
experiments, as indicated, again for the same King model halo 
parameters.  
}
\label{Fig:DAMA_compat_SI_120_3_0.2}
\end{center}
\end{figure}

\begin{figure}[h]
\centering
\begin{tabular}{cc}
\epsfig{file=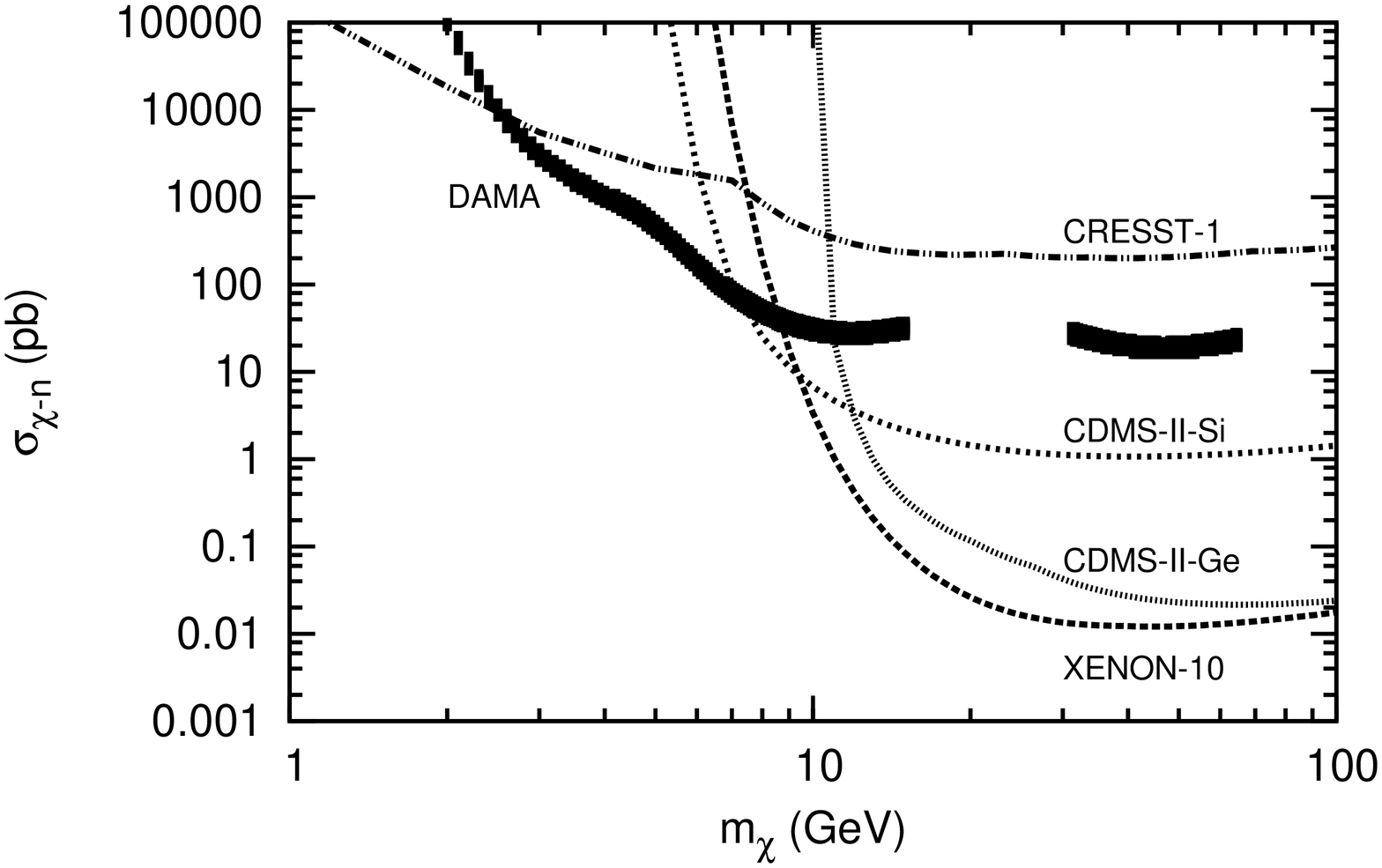,angle=270,width=3.5in}
&
\epsfig{file=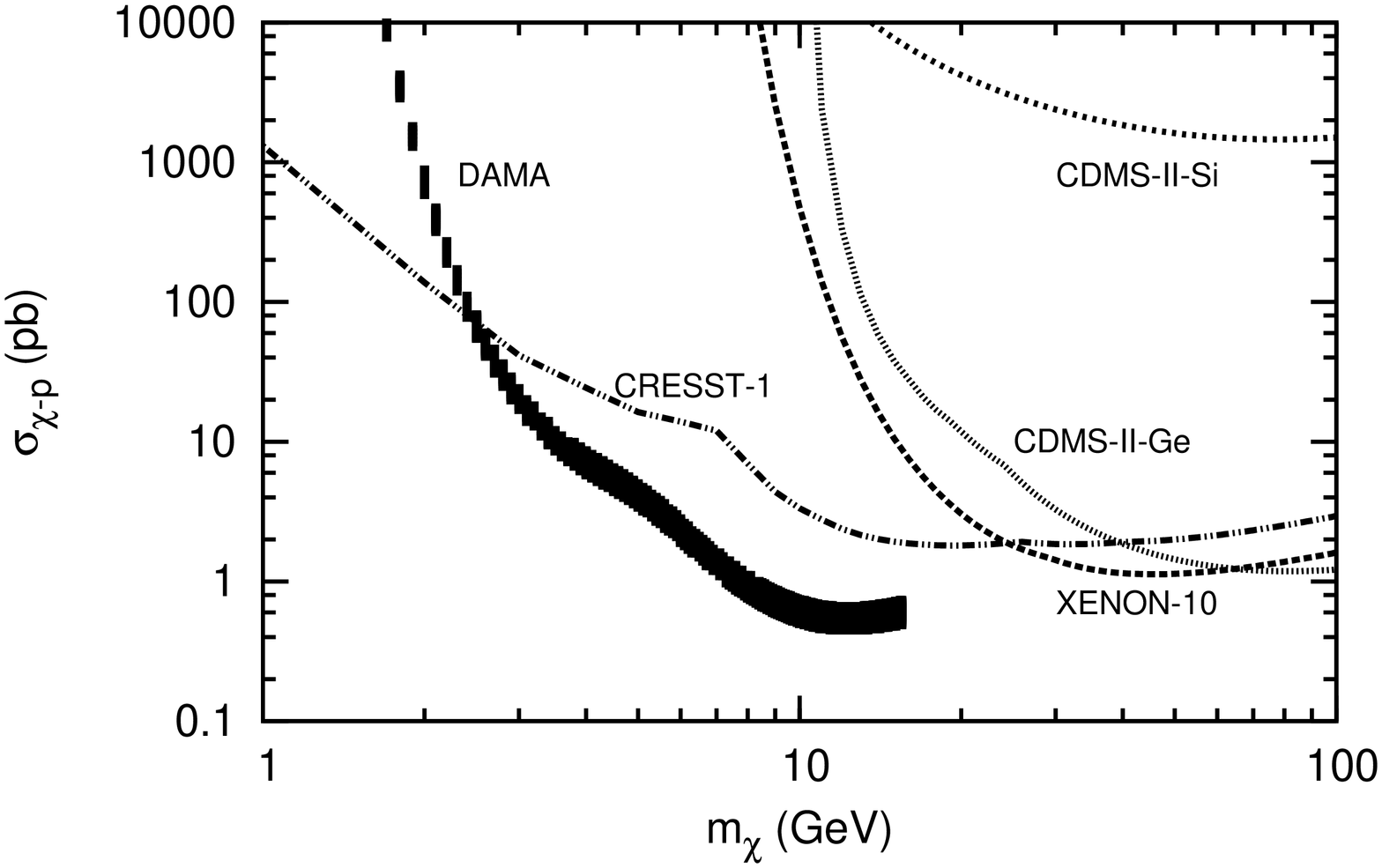,angle=270,width=3.5in}
\end{tabular}
\caption{90\% C.L. allowed region in the WIMP mass vs.~WIMP-neutron 
($a_p=0$) (left panel) and WIMP-proton ($a_n=0$) (right panel) 
spin-dependent cross section plane implied by the DAMA modulation
signal, for our self-consistent truncated isothermal (King) model of
Milky Way's DM halo with parameters $r_t=120\kpc$, $\sigma=300\kmps$ and
$\rhodmsun=0.2\gev/\cm^3$. Also shown are 90\% C.L. upper limits on 
the cross sections as a function of WIMP mass as implied by the null 
results of other experiments, again for the same King model halo 
parameters.
}
\label{Fig:DAMA_compat_SD_120_3_0.2}
\end{figure}

\noindent
Figures \ref{Fig:exclusionplot_CDMS} --  
\ref{Fig:DAMA_compat_SD_120_3_0.2} show our main results in terms of 
the constraints on the relevant WIMP-nucleon cross sections as a 
function of the WIMP mass, as implied by the results of the direct 
detection experiments. In all these Figures, the 
regions above the curves are excluded at the 90\% C.L.~by the 
results of the respective experiments. 

\paras 
Figure \ref{Fig:exclusionplot_CDMS} shows our 90\% C.L.~upper limits on 
WIMP-nucleon spin-independent cross section as a function of WIMP mass 
as implied by the latest CDMS-II results~\cite{cdms-II_Ge_PRL_10}. For 
our halo model with $\rhodmsun=0.3\gev/\cm^3$ (with other halo 
parameters fixed by fitting to the rotation curve data), the lowest  
upper limit on the WIMP-nucleon spin-independent cross section 
is $3.4\times10^{-8}\pb$ at $\mchi=69\gev$ --- coincidentally almost 
identical to the value of $3.8\times10^{-8}\pb$ quoted in 
Ref.~\cite{cdms-II_Ge_PRL_10} which uses SHM in their analysis 
(presumably with a chosen value of the Galactic escape speed of 
$600\kmps$~\cite{Lewin_Smith_96}). However, 
for our self-consistent halo model that gives the best fit to the new 
rotation curve data --- 
this model has $\rhodmsun=0.2\gev/\cm^3$ --- the above upper limit on 
the SI cross section changes to $\sim 5.3\times10^{-8}\pb$ at 
$\mchi=71\gev$. The upper limit value of the cross section 
approximately scales inversely with the value of $\rhodmsun$.  

\paras
Figures \ref{Fig:DAMA_compat_SI_120_3_0.2} and 
\ref{Fig:DAMA_compat_SD_120_3_0.2} show the results of our study of the 
compatibility of the DAMA experiment's claimed positive result (based on 
their original 2-bin annual modulation amplitude data) with the null 
results of various other experiments, for the cases of 
WIMP-nucleon spin-independent and spin-dependent cross sections, 
respectively. 
For the spin-independent case, we find that there is a range of 
small WIMP masses, $2.6 \lsim\mchi \lsim 10.8\gev$, within which 
DAMA's claimed modulation signal is consistent with the null results of 
other experiments. The allowed WIMP-nucleon SI cross section varies from 
$\sim 1.0\times10^{-2}\pb$ at the lower end of the WIMP mass range to 
$\sim 1.2\times10^{-5}\pb$ at the upper mass end.  
For the spin-dependent case, the DAMA-compatible mass range for 
WIMP-neutron interaction (i.e., $a_p=0$) is $2.5 \lsim\mchi \lsim 
7.4\gev$, with the corresponding WIMP-neutron interaction cross section 
range of $9.5\times10^{3}$ -- $42.0$ pb. For the case of WIMP-proton 
interaction (with $a_n=0$), the corresponding WIMP mass and cross 
section ranges are $2.3 \lsim\mchi \lsim 16\gev$ and 79.0 -- 0.45 pb, 
respectively. 

\paras
We emphasize that the DAMA-compatible ranges of WIMP mass and cross 
sections obtained above are based on our simple analysis procedure using 
the original 2-bin DAMA annual modulation amplitude data, and are 
comparable to those obtained earlier 
within the context of the SHM, e.g., in Ref.~\cite{Savage_etal_08} 
(their 2-bin analysis results). More rigorous analyses 
using DAMA bins over smaller intervals, as done, e.g., in 
\cite{Savage_etal_08}, should be able to place more restrictive 
constraints on the DAMA-compatible ranges of WIMP mass and cross section 
within the context of our halo model. This will be taken up in a work 
now in progress.  

\section{Summary and Conclusions
\label{sec:summary_conclusions}}
\noindent
The standard halo model (SHM) of the dark matter halo of the Galaxy, 
customarily used in the analysis of the results of WIMP direct 
detection experiments, is not suitable for describing a finite-size 
system such as the Galaxy. In the SHM, the DM halo of the Galaxy is 
described as an isothermal sphere with a Maxwellian velocity 
distribution of its constituent particles. The isothermal sphere 
solution of the steady-state collisionless Boltzmann equation is 
actually infinite 
in extent and has a divergent total mass with the mass contained inside 
a radius $r$ increasing linearly with $r$~\cite{Binney_Tremaine}. In the 
literature on direct detection of WIMP dark matter particles, it is a 
common practice to truncate the Maxwellian speed distribution assumed in 
the SHM at a chosen value of the local escape speed of the Galaxy. This, 
however, is not a self-consistent procedure because the resulting 
truncated speed distribution does not in general satisfy the 
steady-state collisionless Boltzmann equation which a system of 
collisionless particles such as WIMPs should. Moreover, the results 
of analysis of the outcome of a direct 
detection experiment would depend on the value of the escape speed 
chosen, which is quite uncertain (see, e.g., 
Ref.~\cite{Smith_etal_MNRAS_07}). 

\paras 
In this paper we have presented a self-consistent model of the 
finite-size dark halo of the Galaxy with its phase space distribution 
function described by a truncated isothermal model (``King" model). We 
have also included the gravitational influence of the observed visible 
matter on the structure of the DM halo in a self-consistent manner. 
In this model the velocity distribution function of the WIMPs 
constituting the halo is non-Maxwellian, with a cut-off at a maximum 
velocity that is self-consistently determined by the model itself.  
We have determined the parameters of our halo model by a fit to a 
recently determined circular rotation curve of the Galaxy that extends 
up to a Galactocentric distance of 
$\sim60\kpc$~\cite{rc_60kpc_Xue_etal_08}. A noticeable feature of this 
rotation curve is that it declines with the Galactocentric radius beyond 
$\sim 10\kpc$. This imposes strong constraints on the 
(truncation) radius, density and mass of the DM halo and hence on the 
total mass of the Galaxy. 
Specifically, for values of local DM density in the range $\rhodmsun = 
0.2$ -- $0.4\gev/\cm^3$, reasonable fits to the rotation curve data 
require the truncation radius of the halo to be in the 
corresponding range $r_t\approx$ 120 -- 80 kpc, in all cases restricting 
the total mass of the Galaxy (including its DM halo) to relatively low 
values of $M_{\rm Galaxy}\lsim 3\times10^{11}\msun$. Note, however, 
that the rotation curve estimated in Ref.~\cite{rc_60kpc_Xue_etal_08}, 
which is used in determining the parameters of our halo model,  
was derived assuming a value of $\approx 220\kmps$ for the local 
standard of rest. On the other hand, as suggested 
in Ref.~\cite{new_rot_speed}, the value of the local standard of rest 
may have to be revised upward to a value of $\vcsun\approx 250\kmps$. 
This would require appropriate scaling up of the rotation curve data of 
Ref.~\cite{rc_60kpc_Xue_etal_08}, which in turn will require 
an appropriately higher value of $\rhodmsun$ and a correspondingly 
higher value of the total mass of the Galaxy. 

\paras 
Interestingly, we find that the upper limits on the relevant
WIMP-nucleon interaction cross section implied by the null 
results of the direct detection experiments are primarily determined by 
the chosen value of the local DM density $\rhodmsun$ --- scaling roughly
inversely with the value of $\rhodmsun$ --- and are relatively less 
sensitive to the other parameters of the model such as the truncation 
radius and total mass of the halo. For our best-fit self-consistent 
lowered isothermal halo model with parameter values 
$\rhodmsun=0.2\gev/\cm^3$ and 
$r_t=120\kpc$, that provides the best fit to the rotation curve data of 
Ref.~\cite{rc_60kpc_Xue_etal_08}, the null result of the CDMS-II 
experiment~\cite{cdms-II_Ge_PRL_10}, for example, gives a 90\% C.L. 
upper limit on the WIMP-nucleon spin-independent (SI) interaction cross 
section of $\sim 5.3\times10^{-8}\pb$ at a WIMP mass of 71 GeV. 
  
\paras
Concerning the issue of the compatability of the claimed 
positive signal reported by the DAMA collaboration with the null results 
from the other experiments, we find, within the context of a simple 
analysis procedure using the original 2-bin DAMA annual modulation data, 
that there exist regions of the WIMP mass vs.~WIMP-nucleon cross section 
with small WIMP masses typically in the range $\sim$ 2 -- 16 GeV, within 
which the DAMA's claimed annual modulation
signal is consistent with the null results of other experiments. 
While this strengthens the possibility of a low mass WIMP as a DM 
candidate --- a possibility also indicated by several earlier analyses 
done using the SHM --- a more rigorous analysis using DAMA bins over 
smaller intervals must be performed before a more definitive 
conclusion can be reached in this regard. Very recently, the CoGeNT 
collaboration~\cite{cogent_arxiv_1002.4703} has also reported an excess 
of low (nuclear recoil) energy events, which may also be pointing 
towards a low-mass WIMP DM candidate. We shall discuss these issues  
within the context of our self-consistent model of the Galaxy's dark 
halo in a work now in progress.      

\paras
\section*{Acknowledgments}
We thank Susmita Kundu for discussions.

\end{document}